\documentclass[prb,aps,twocolumn,amsfonts,showpacs,superscriptaddress]{revtex4-1}

\usepackage{amsmath,amssymb,bm} 
\usepackage{graphicx}  
\usepackage{color}
\usepackage{subfigure}    
\usepackage[colorlinks=true]{hyperref}  
\hypersetup{
    bookmarks=true,         
    unicode=false,          
    pdftoolbar=true,        
    pdfmenubar=true,        
    pdffitwindow=false,     
    pdfstartview={FitH},    
    pdftitle={Shear viscosity at the Ising-nematic quantum critical point in two dimensional metals},    
    pdfauthor={Aavishkar A. Patel, Andreas Eberlein, Subir Sachdev},     
    pdfsubject={},   
    pdfcreator={},   
    pdfproducer={}, 
    pdfkeywords={} {} {}, 
    pdfnewwindow=true,      
    colorlinks=true,       
    linkcolor=magenta, 
    citecolor=blue,        
    filecolor=magenta,      
    urlcolor=blue           
} 
\usepackage{amsfonts}
\usepackage{upgreek}
\usepackage{slashed}
\usepackage{latexsym}

\usepackage{todonotes}

\newcommand{\beq}{\begin{equation}}
\newcommand{\eeq}{\end{equation}}
\def\bea{\begin{eqnarray}}
\def\eea{\end{eqnarray}}

\newcommand{\bs}{\boldsymbol}
\newcommand{\intd}[2]{\int\frac{d^{#2} #1}{(2\pi)^{#2}}}
\newcommand{\intf}[1]{\int\frac{d #1}{2\pi}}
\newcommand{\intm}[1]{\int_{#1}}
\newcommand{\intdef}[3]{\int_{#2}^{#3}\negthickspace\frac{d #1}{%
2\pi}}
\newcommand{\intdefnopi}[3]{\int_{#2}^{#3}\negthickspace d #1}

\begin{document}

\title{Shear viscosity at the Ising-nematic quantum critical point\\ in two dimensional metals}

\author{Andreas Eberlein}
\affiliation{Department of Physics, Harvard University, Cambridge, 
Massachusetts 02138, USA}

\author{Aavishkar A. Patel}
\affiliation{Department of Physics, Harvard University, Cambridge, 
Massachusetts 02138, USA}

\author{Subir Sachdev}

\affiliation{Department of Physics, Harvard University, Cambridge, 
Massachusetts 02138, USA}
\affiliation{Perimeter Institute for Theoretical Physics, Waterloo, Ontario, Canada N2L 2Y5}

\date{\today}

\begin{abstract}%
In an isotropic strongly interacting quantum liquid without quasiparticles, general scaling arguments imply that the dimensionless ratio
$(k_B /\hbar)\, \eta/s$, where $\eta$ is the shear viscosity and $s$ is the entropy density, is 
a universal number. We compute the shear viscosity of the Ising-nematic critical
point of metals in spatial dimension $d=2$ by an expansion below $d=5/2$. 
The anisotropy associated with directions parallel and normal to the Fermi surface leads to a violation of the scaling
expectations: $\eta$ scales in the same manner as a chiral conductivity, and the ratio  $\eta/s$ diverges at low temperature ($T$) 
as $T^{-2/z}$,
where $z$ is the dynamic critical exponent for fermionic excitations dispersing normal to the Fermi surface.
\end{abstract}


\maketitle

\section{Introduction}
\label{sec:intro}

Recent experiments on graphene \cite{Geim16,Kim16} and PdCoO$_2$ \cite{APM16} have displayed remarkable
evidence for nearly-momentum-conserving hydrodynamic flow of the electron liquid. In clean Fermi liquids,
hydrodynamic flow requires very clean samples with weak umklapp scattering so that electron-electron collisions
lead to thermalization before there is significant momentum lost to the crystal \cite{APM16,Gurzhi,Molen95,AKS10}. However, rapid thermalization and hydrodynamics
are natural properties of quantum critical systems \cite{Damle97}
and strange metals \cite{HKMS,HMPS14,PS14,LS15}, 
and their consequences should be visible even in moderately clean samples. Graphene was proposed as a strange metal
in which ill-defined quasiparticles lead to hydrodynamic flow at intermediate temperatures \cite{MMSS08,FSMS08,MFS08,FA09,MSF09,Mendoza11,Polini14,Vignale15,Polini15,Levitov16,ALJC15,Levitov16b}:
the experiments also display evidence \cite{Polini15,Levitov16,ALJC15,Geim16,Kim16} for the viscous
drag of such flow. There have also been studies of viscous flow in high energy physics \cite{PSS01,KSS05,KKT08,HSS12} and ultracold atoms \cite{Thomas11,Taylor2010,Enss2011,AK14,MBTS16}.

These experimental advances indicate that the time is ripe for exploring hydrodynamic electron flow in the ubiquitous 
strange metal regimes of the cuprates or the pnictides. These are metals without quasiparticle excitations,
and so should exhibit hydrodynamic flow when impurities are dilute.
We note the indirect evidence for such behavior in the photoemission experiments
of Rameau {\em et al.\/} \cite{Johnson14}.
To this end, here we examine the simplest model which realizes
a metallic state without quasiparticles in two spatial dimensions, and compute its shear viscosity, $\eta$.
We will study the quantum critical point (QCP) for the onset of Ising nematic order \cite{CHWM00,OKF01,MRA03}
using its continuum field theoretic formulation using patches on the Fermi 
surface \cite{Metlitski2010a, Dalidovich2013}. 

General scaling arguments (reviewed below) for a spatially isotropic system
imply that $\eta$ should scale in the same manner as the entropy density, $s$; so 
\beq 
\eta/s \sim \hbar/k_B, \label{etaos}
\eeq
where the r.h.s. restores dimensions, and the prefactor is expected to be of order unity. (In $d=2$ hydrodynamic long time tails
can lead to logarithmic corrections to $\eta$ \cite{LTT} which we ignore here, as we find much larger corrections). This is also the expectation
from holographic studies of critical quantum liquids \cite{PSS01,KSS05,Nabil09,Raychowdhury15,Kiritsis15,KuangWu15,KMN16}. 
A relationship of the form (\ref{etaos}) appeared in string-theoretic realizations of strongly-coupled field theories \cite{KSS05},
and has been widely used as a diagnostic of strongly-coupled non-quasiparticle dynamics in the quark-gluon plasma \cite{PSS01,KSS05,KKT08,HSS12}.

\begin{figure}
	\includegraphics[width=0.6\linewidth]{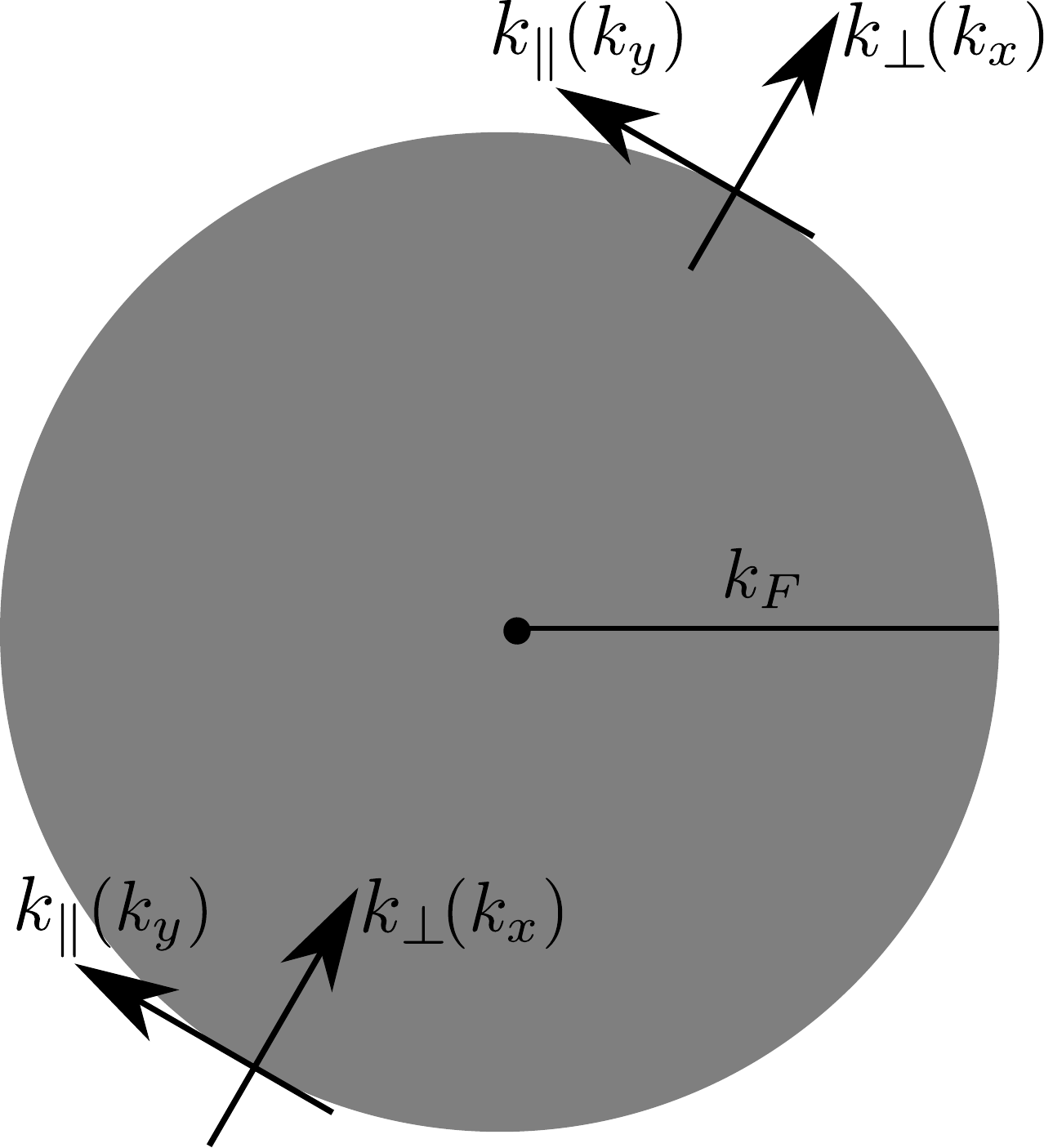}
	\caption{Fermi surface and definition of the momentum components 
parallel ($k_\parallel$) and perpendicular ($k_\perp$) to the Fermi surface at 
the two Fermi surface patches in which the low-energy field theory is defined.}
	\label{fig:FermiSurface}
\end{figure}

Our main result is that Eq.~(\ref{etaos}) does not apply to many of the models of electronic strange metals without quasiparticles. 
Even without long-lived quasiparticles, such models have a Fermi surface at $T=0$,  which defines momenta
with singular low energy excitations; more precisely, the Fermi surface is the locus of points at which the inverse Green's function 
vanishes. Although the metal is globally isotropic, the excitations in the vicinity of a particular point
on the Fermi surface have a highly anisotropic structure, as shown in 
Fig.~\ref{fig:FermiSurface}: excitations at a momentum $k_\perp$ perpendicular 
to the Fermi surface
have a typical energy $k_\perp^z$, while excitations at a momentum $k_\parallel$ parallel to the Fermi surface have a typical
energy $k_\parallel^{2z}$; here $z$ is the dynamic critical exponent. 
In the present 
paper, we will show that the dispersion of the excitations parallel to the Fermi 
surface plays a more fundamental role in determining the value of the shear 
viscosity $\eta$. As a result Eq.~(\ref{etaos}) does not apply, and we 
find instead a divergence as $T \rightarrow 0$,
\beq
\eta/s \sim T^{-2/z}. \label{etaos2}
\eeq
This surprising violation of (\ref{etaos}) in an isotropic system can be traced directly to the presence of a Fermi surface. 
Our result implies that holographic duals of strange metals \cite{Liu09,Zaanen09,Kiritsis10,Trivedi10,Ogawa11,HSS11} 
do not fully capture the Fermi surface structure. Instead, it appears that bulk quantum gravity corrections
will be required to resurrect the Fermi surface in the holographic theories \cite{Polchinski:2012nh,Faulkner:2012gt,Sachdev:2012tj}, and to obtain the result corresponding
to Eq.~(\ref{etaos2}).

Section~\ref{sec:scaling} will present a review of scaling arguments which usually apply the conventional relation
in Eq.~\eqref{etaos}. The dimensionally extended field theory for the quantum critical point will be presented in 
Section~\ref{sec:field}. We will use this field theory to compute the `optical' shear viscosity ({\em i.e.\/}
the viscosity at frequencies $\omega \gg T$) in Section~\ref{sec:viscosity}. We will then examine the usual DC viscosity
(at frequencies $\omega \ll T$) in Section~\ref{sec:dcviscosity}.

\section{Scaling arguments}
\label{sec:scaling}

In studies so far of the thermodynamic and transport
properties of strange metals, the anisotropy of the Fermi surface has had a 
specific consequence \cite{Eberlein2016}: the entropy density, and the electrical and thermal
conductivities are dominated by the energy dispersion perpendicular to the Fermi surface, while the direction parallel
to the Fermi surface mostly acts as a label which counts the total density of perpendicular excitations. 
Consequently, in scaling arguments 
we find a violation of hyperscaling: this is the property in which the entropy density of a $d$ dimensional system scales as if 
it is in $d-\theta$
dimensions, with $\theta$ the violation of hyperscaling exponent.
For a Fermi surface $\theta=d-1$, because only the dispersion perpendicular to 
each point on the Fermi surface is important in the computation of the entropy. 
Recent work has shown~\cite{Eberlein2016} that similar arguments also correctly 
determine the electrical conductivity and entropy density. 

We now review the general scaling arguments for the universality of $\eta/s$.
The entropy density invariably scales as a density, and so has scaling dimension $d$.
From the arguments just presented above, with the violation of 
hyperscaling in the presence of a Fermi surface, the entropy density $s$ 
should have scaling dimension $d-\theta$, and so
\beq
s \sim T^{(d-\theta)/z} ; \label{stheta}
\eeq
this was confirmed by computations in \cite{Eberlein2016}.
Similar arguments apply to the optical conductivity $\sigma_Q (\omega)$, where $\omega$ is a frequency; 
naively, the conductivity has scaling dimension $d-2$, and so we can expect that in the presence of a Fermi surface, the dimension
will be $d-\theta - 2$. The computation in \cite{Eberlein2016} shows that this 
is indeed the case, and we have 
\beq
\sigma_Q \sim T^{(d-2-\theta)/z} \Upsilon ( \omega/T) , \label{sigmaQtheta}
\eeq
where $\Upsilon$ is a scaling function.

In an isotropic system that obeys hyperscaling, we can read off the 
scaling dimension of the stress tensor from its 
definition~\cite{schwartz2014quantum},
\begin{equation}
	T_{\mu\nu} = \sum_n \Bigl(\frac{\delta \mathcal{L}}{\delta(\partial_\mu \zeta_n)} 
\partial_\nu \zeta_n - \partial_\mu \frac{\delta \mathcal{L}}{\delta(\partial^2_\mu 
\zeta_n)} \partial_\nu \zeta_n \Bigr) - \delta_{\mu\nu} \mathcal{L},
\end{equation}
where $\zeta_n$ denotes all the fields in the theory and $\mathcal{L}$ the 
Lagrangian density.
It follows that the  spatial components have the same scaling dimension as the 
Lagrangian density,
 $[T_{ij}]  = d + z$, and that the mixed temporal-spatial components have scaling dimension 
$[T_{0i}] = d + 1$. Inserting these scaling dimensions into the Euler equation,
\begin{equation}
	\partial_t p_\alpha = \partial_\beta T_{\alpha\beta},
\end{equation}
where $p_\alpha$ are the components of the momentum operator, yields consistent 
results. We thus obtain \begin{equation}
	[T_{xy}] = d + z
\end{equation}
in the presence of hyperscaling. Kubo's formula for the frequency-dependent 
shear viscosity reads~\cite{Taylor2010,Enss2011},
\begin{equation}
	\operatorname{Re} \eta(\omega) = \lim_{\bs q \rightarrow 0} \omega^{-1} 
\operatorname{Im} \chi_{T_{xy} T_{xy}}(\omega, \bs q),
\label{eq:Kubo}
\end{equation}
where
\begin{equation}
	\chi_{T_{xy} T_{xy}}(i\omega_n, \bs q) = \langle T_{xy}(i\omega_n, \bs q) 
T_{xy}(-i\omega_n, -\bs q)\rangle
\end{equation}
is the autocorrelation function of the $xy$-component of the stress 
tensor $T$. Its scaling dimension is
\begin{equation}
	[\eta] = -z - d - z + 2 [T_{xy}] = d
\end{equation}
and the d.c. shear viscosity is given by $\eta = \lim_{\omega\rightarrow 0} 
\eta(\omega)$. This is the same scaling dimension as for the entropy density 
above. With the violation of hyperscaling in the presence of a Fermi surface, 
the examples of the entropy density and the optical conductivity above suggest 
that $\eta$ should scale just like
$s$ in Eq.~(\ref{stheta}), and hence Eq.~(\ref{etaos}) should apply. Our computations in this paper
show that this is not true, and the Fermi surface leads to behavior genuinely different both from naive
scaling assumptions, and from holographic examples: the $T$ dependence of $\eta$ is such that
Eq.~(\ref{etaos2}) holds.

\section{Field theory}
\label{sec:field}
 
We now recall the field theory which
allow us to formulate a systematic and controlled 
renormalization group analysis using a convenient dimensional regularization
method. Moreover, this method fully preserves a two-dimensional Fermi surface 
with anisotropic dispersion in the vicinity of every point on the Fermi surface, 
and these features are crucial for our results. We will discuss the field 
theory for the Ising-nematic critical point, but similar field theories and 
results also apply to the problem of a Fermi surface coupled to a gauge field, 
or to other long-wavelength order parameters \cite{Metlitski2010a}.

\begin{widetext}
We consider a theory of fermions, $\psi$, in $(2+1)$ dimensions which are coupled to a 
critical boson, $\Phi$,
\begin{equation}
\begin{split}
	S(\bar\psi,\psi,\Phi) = &\sum_{s = \pm} \sum_{j = 1}^N \intd{k}{3} 
\tilde\psi^\dagger_{s j}(k) (i k_0 + s k_x + k_y^2) \tilde \psi_{sj}(k) + 
\frac{1}{2} \intd{k}{3} (k_0^2 + k_x^2 + k_y^2) \Phi(-k) \Phi(k)\\
	& + \frac{e}{\sqrt{N}} \sum_{s = \pm} \sum_{j = 1}^N \intd{k}{3} \intd{q}{3} 
\lambda_s \Phi(q) \tilde\psi^\dagger_{s j}(k+q) \tilde \psi_{s j}(k),
\end{split}
\label{eq:Action2D}
\end{equation}
where $e$ is the fermion-boson coupling constant, $s = \pm 1$ labels the two 
Fermi surface patches, $N$ is the number of fermionic flavors and $\lambda_s$ 
equals 1 ($s$) for the Ising-nematic critical point (fermions coupled to a 
$U(1)$ gauge field). This model has been studied by many authors, including 
Refs.~\onlinecite{Metlitski2010a, Dalidovich2013}. In the following, we restrict 
ourselves to the Ising-nematic critical point and set $\lambda_s = 1$. 

This model can be studied in a controlled way using the dimensional 
regularization scheme proposed by Dalidovich and Lee~\cite{Dalidovich2013}. 
Increasing the codimension of the Fermi surface by introducing 
auxiliary time-like directions, the dimensionally regularized action in $(d+1)$ 
dimensions reads
\begin{equation}
\begin{split}
	S(\bar\psi,\psi,\Phi) = & \sum_{j = 1}^N \intd{k}{d+1} \bar\psi_j(k) [i\bs 
\Gamma \cdot \bs K + i \gamma_x \delta_k] \psi_j(k) + \frac{1}{2} \intd{q}{d+1} 
[\bs Q^2 + q_x^2 + q_y^2] \Phi(-q) \Phi(q)\\
& + \frac{i e}{\sqrt{N}} \sqrt{d-1} \sum_{j = 1}^N \intd{k}{d+1} \intd{q}{d+1} 
\Phi(q) \bar\psi_{j}(k+q) \gamma_x \psi_j(k),
\end{split}
\end{equation}
where $\bs K = (k_0, k_1, \ldots, k_{d-2})$ collects the physical and $(d-2)$ 
auxiliary frequency variables. We introduced the spinor notation
\begin{align}
	\psi_j(k) &= \begin{pmatrix}
	             	\tilde \psi_{+,j}(k), \tilde \psi^\dagger_{-,j}(-k)
	             \end{pmatrix}^T		&%
	\bar\psi_j(k) &= \psi^\dagger_j(k) \gamma_0
\end{align}
and defined the gamma matrices as $\gamma_0 = \sigma_y$ and $\gamma_x = 
\sigma_x$ for the spatial and as $\bs \Gamma = (\gamma_0, \gamma_1, 
\ldots, \gamma_{d-2})$ for the time-like directions.
Within a patch, we choose $k_x$ ($k_y$) perpendicular 
(parallel) to the Fermi surface, as shown in Fig.~\ref{fig:FermiSurface}. The 
dispersion in the spatial plane containing 
the Fermi surface is $\delta_k$, while the full dispersion is $\varepsilon_k$ 
with
\beq
\delta_k = k_x + 
\sqrt{d-1} k_y^2 \quad , \quad \varepsilon_k = \left( \delta_k^2 + 
\sum_{i=1}^{d-2} k_i^2 \right)^{1/2}.
\eeq
Note the line of zero energy excitations in the plane $k_i=0$ which represents a patch on the Fermi surface in 
Fig.~\ref{fig:FermiSurface}, and the relativistic dispersion along the $k_i$ 
directions.

Rescaling momenta as
\begin{align}
	\bs K &= b^{-1} \bs K'		&		k_x &= b^{-1} k_x'		& k_y = b^{-1/2} 
k_y',
\end{align}
the fermionic quadratic part of the action and the contribution $\sim q_y^2$ 
in the bosonic quadratic part of the action are invariant under rescaling if 
fields are scaled as
\begin{align}
	\psi_j(k) &= b^{d/2 + 3/4} \psi_j'(k')	&		\Phi(k) &= b^{d/2 + 3/4} \Phi'(k')
\end{align}
The terms proportional to $\bs Q^2$ and $q_x^2$ in the bosonic quadratic 
part are irrelevant under this rescaling. The coupling scales as
\begin{equation}
	e' = e b^{\frac{1}{2}(5/2-d)},
\end{equation}
identifying $d = 5/2$ as the upper critical dimension. It is 
irrelevant for $d > 5/2$ and relevant for $d < 5/2$. This allows to access 
non-Fermi liquid physics perturbatively by using $\epsilon = 5/2 - d$ as 
expansion parameter.

Keeping only marginal terms, the ansatz for the local field theory reads
\begin{equation}
\begin{split}
	S(\bar\psi,\psi,\Phi) = &\sum_{j = 1}^N \intd{k}{d+1} \bar\psi_j(k) [i 
\bs\Gamma \cdot \bs K + i \gamma_x\delta_k] \psi_j(k) + \frac{1}{2} 
\intd{q}{d+1} q_y^2 \Phi(-q) \Phi(q)\\
& + \frac{i e \mu^{\epsilon/2}}{\sqrt{N}} \sqrt{d-1} \sum_{j = 1}^N 
\intd{k}{d+1} 
\intd{q}{d+1} \Phi(q) \bar\psi_j(k+q) \gamma_x \psi_j(k),
\end{split}
\label{eq:action}
\end{equation}
where we introduced the momentum scale $\mu$ in order to make the coupling $e$
dimensionless. Perturbative corrections to this action at one-loop level 
reintroduce dynamics for the bosonic field. The $\epsilon=5/2-d$ expansion 
allows us to make a renormalized perturbative computation in the dimensionless 
coupling $e$. Note that this is not equivalent to a simple $1/N$ expansion, 
which breaks down at the Ising-nematic QCP~\cite{Metlitski2010a}, and that the 
expansion parameter is $\frac{e^{4/3}}{N}$~\cite{Dalidovich2013}.
\end{widetext}

\section{Optical shear viscosity}
\label{sec:viscosity}

In the following, we focus on the `optical' shear viscosity, evaluated
at frequencies $\omega \gg T$. 
Its evaluation is simpler than that for the d.c.~viscosity, $\omega \ll T$, 
which will be considered in Section~\ref{sec:dcviscosity}.  

For the 
Ising-nematic QCP the $xy$-component of the stress tensor is proportional to 
the $y$-component of the `chiral' current operator,
\begin{equation}
\begin{split}
	T_{xy}(q) &= i \sum_{j = 1}^N \intm{k} \bigl(k_y + \frac{q_y}{2}\bigr) 
\bar\psi_j(k+q) \gamma_{x} \psi_j(k)\\
	&= \frac{1}{2\sqrt{d-1}} J_y(q).
\end{split}
\end{equation}
where $\intm{k} = \intf{k_x}{}\intf{k_y}{}\intd{K}{d-1}$. Note that the $x$- and $y$-components 
of the chiral current contain the same gamma matrix.

\begin{figure}
	\centering
	\subfigure[]{\includegraphics[width=0.4\linewidth]{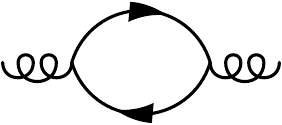} \label{fig:1L}}
	\subfigure[]{\includegraphics[width=0.4\linewidth]{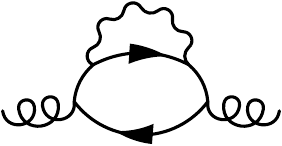} \label{fig:2LSE}}
	\subfigure[]{\includegraphics[width=0.4\linewidth]{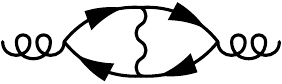} \label{fig:2LVC}}
	\caption{Feynman diagrams yielding the renormalization of the scaling 
behavior of the viscosity at lowest order in $\epsilon$: (a) One-loop 
contribution, (b) self-energy correction and (c) vertex correction. Lines 
represent fermionic propagators, wiggly lines bosonic propagators and curly 
lines the stress tensor.}
	\label{fig:FeynmanDiagrams}
\end{figure}
The Feynman diagrams describing the renormalization of 
the scaling behavior of the viscosity at lowest order in $\epsilon$ are shown 
in Fig.~\ref{fig:FeynmanDiagrams}. 

At one-loop level, the stress tensor 
correlator is given by
\begin{align}
&\langle T_{xy}(q) T_{xy}(-q)\rangle_\text{1Loop} = \nonumber \\
&= N \intm{k} (k_y + q_y/2)^2 \operatorname{tr}\bigl(\gamma_x G_0(k+q) \gamma_x 
G_0(k)\bigr)
\end{align}
where
\begin{equation}
	G_0(k) = \frac{\bs \Gamma\cdot \bs K + \gamma_x \delta_k}{i(\bs K^2 + 
\delta_k^2)}
\end{equation}
is the bare fermionic Green's function. Specializing to $q = \omega \bs e_0$,
\begin{equation}
\begin{split}
	& \langle T_{xy} T_{xy} \rangle_\text{1Loop}(i\omega) = \\
	&= -2N \intd{k}{d+1} k_y^2 \frac{\delta_k^2 - \bs K \cdot (\bs K + \bs 
Q)}{(\bs K^2 + \delta_k^2)\bigl((\bs K + \bs Q)^2 + \delta_k^2\bigr)},
\end{split}
\end{equation}
where $\bs Q = \omega \bs e_0$. The further evaluation parallels that of 
the optical conductivity~\cite{Eberlein2016}. Shifting $k_x \rightarrow k_x - 
\sqrt{d-1} k_y^2$ eliminates $k_y$ from the integrand except in the prefactor 
arising from the stress tensor, yielding
\begin{equation}
\begin{split}
& \langle T_{xy} T_{xy}\rangle_\text{1Loop}(i\omega) = \\
& -2 N 
\intdef{k_y}{}{} k_y^2 \intdef{k_x}{}{} 
\intd{K}{d-1} \frac{k_x^2 - \bs K 
\cdot (\bs K + \bs Q)}{(\bs K^2 + k_x^2) ((\bs K + \bs Q)^2 + k_x^2)} \\
	&= -2 N \intdef{k_y}{}{} k_y^2 I_\text{1loop}(\bs Q).
\end{split}
\end{equation}
Introducing Feynman parameters, completing squares in the denominator and 
shifting $\bs K \rightarrow \bs K - (1-x) \bs Q$, we obtain
\begin{equation}
\begin{split}
	& I_\text{1loop}(\bs Q) =\\
	&  \intd{K}{d-1}\intd{p}{} \intdefnopi{x}{0}{1} 
\frac{p^2 - \bs K^2 + x (1-x) 
\bs Q^2}{[\bs K^2 + p^2 + x (1 - x) \bs Q^2]^2}\\
&= \frac{\pi S_{d-1}}{(2\pi)^d} \intdefnopi{k}{0}{\infty} k^{d-2} 
\intdefnopi{x}{0}{1} \frac{ x (1-x) \bs Q^2}{[k^2 + x (1-x) \bs Q^2]^{3/2}}\\
&= \frac{S_{d-1}}{(2\pi)^d} \sqrt{\pi} \Gamma(2 - d/2) 
 \frac{\Gamma(\frac{d-1}{2})\Gamma(d/2)^2}{\Gamma(d)} |\bs Q|^{d-2}
\end{split}
\end{equation}
($S_d=2\pi^{d/2}/\Gamma(d/2)$). For $d = 5/2 - \epsilon$, the one-loop result for the stress tensor 
autocorrelation function thus reads
\begin{equation}
	\langle T_{xy} T_{xy} \rangle_\text{1loop} (i\omega) = -N 
u_{\text{1Loop},\epsilon} \intdef{k_y}{}{} k_y^2 |\omega|^{1/2-\epsilon},
\label{1loop}
\end{equation}
where
\begin{equation}
u_{\text{1Loop},\epsilon} = \frac{2^{\epsilon-1/2} 
\Gamma(\frac{3+2\epsilon}{4}) 
\Gamma(\frac{5-2\epsilon}{4})^2}{\sqrt{\pi}^{5/2-\epsilon} 
\Gamma(\frac{5-2\epsilon}{2})}.
\end{equation}
The momentum parallel to the Fermi surface, $k_y$, does not scale due to the 
emergent rotational symmetry~\cite{Metlitski2010a} of the 
low-energy field theory. The latter restricts the momentum dependence of the 
fermionic and bosonic propagator to $G(\bs K, k_x, k_y) = G(\bs K, \delta_k)$ and 
$D(\bs Q, q_x, q_y) = D(\bs Q, q_y)$, respectively, which allows to eliminate $k_y$ 
from the integrand by shifting $k_x$. The $k_y$-integral is cut off by the Fermi 
surface curvature. As a consequence, the result~\eqref{1loop} differs from the 
current-current correlation function only by the fact that $\intm{k_y} k_y^2$ 
appears instead of $\intm{k_y}$~\cite{Eberlein2016}. Importantly, both results 
have the same 
dependence on frequency. 

The two-loop self-energy correction to the optical viscosity is given by
\begin{equation}
\begin{split}
	\langle T_{xy} T_{xy}&\rangle_\text{SE} (i\omega) = 2 N \intm{k} (k_y + 
q_y/2)^2\\
	&\times \operatorname{tr} \Bigl(\gamma_x G_0(k+q) \gamma_x G_0(k) 
\Sigma_1(k) G_0(k)\Bigr),
\end{split}
\end{equation}
where
\begin{equation}
	\Sigma_1(k) = -i(\bs\Gamma\cdot \bs Q) \frac{e^{4/3}}{N} 
\Bigl(\frac{\mu}{|\bs Q|}\Bigr)^{2\epsilon/3} u_{\Sigma,0} \epsilon^{-1} + 
\mathcal O(\epsilon^0)
\end{equation}
is the one-loop fermionic self-energy~\cite{Dalidovich2013} ($u_{\Sigma,0}=(2\cdot6^{1/3}\pi)^{-1}$). We obtain
\begin{equation}
	\langle T_{xy} T_{xy}\rangle_\text{SE} (i\omega) = e^{4/3} \epsilon^{-1} 
\intm{k_y} k_y^2 |\omega|^{1/2-\epsilon} 
\Bigl(\frac{\mu}{|\omega|}\Bigr)^{2\epsilon/3} a_{\Sigma,0},
\end{equation}
where $a_{\Sigma,0} = u_{\text{1Loop},0} u_{\Sigma,0}$, after evaluation of the 
integrals as described in Appendix~\ref{app:twoloop}. The dependence on 
frequency is the same as in the self-energy correction to the current-current 
correlation function~\cite{Eberlein2016}. 

The two-loop vertex correction is 
given by
\begin{equation}
	\begin{split}
		\langle T_{xy} T_{xy}&\rangle_\text{VC} (i\omega) = -i N \intm{k} (k_y + 
q_y/2) \\
	&\times \operatorname{tr} \Bigl(\gamma_x G_0(k+q) \Gamma_{xy,1}(k, q) 
G_0(k)\Bigr),
	\end{split}
\end{equation}
where $\Gamma_{xy,1}$ is the one-loop correction to the stress-tensor. Ward 
identities due to the conservation of the chiral current imply that the vertex 
correction to the stress tensor correlation function does not have a pole in 
$\epsilon^{-1}$, as for the optical 
conductivity~\cite{Dalidovich2013,Eberlein2016}. At lowest order in $\epsilon$, 
we thus obtain
\begin{equation}
\begin{split}
	\langle T_{xy} T_{xy}&\rangle(i\omega) = -N \intm{k_y} k_y^2 
u_\text{1Loop,$0$} \\
	&\times |\omega|^{1/2-\epsilon} \Bigl\{1 - 
\frac{e^{4/3}}{N\epsilon} \Bigl(\frac{\mu}{|\omega|}\Bigr)^{2\epsilon/3} 
u_{\Sigma,0}\Bigr\} + \ldots
\end{split}
\label{eq:TxyTxy}
\end{equation}
for the correlator of the stress tensor. Evaluation of the coupling 
${e^{4/3}}/{N}$ at the fixed point using the $\beta$ function in $\mathcal 
O(\epsilon)$~\cite{Dalidovich2013},
\begin{equation}
	\Bigl(\frac{e^{4/3}}{N}\Bigr)^\ast = u_{\Sigma,0}^{-1} 
\epsilon,
\end{equation}
and resummation of the frequency dependence yields $\langle T_{xy} 
T_{xy}\rangle(i\omega) \sim |\omega|^{1/2-\epsilon/3}$ for the correlator and 
\begin{equation}
	\eta(\omega) \sim \omega^{-1/2 - \epsilon / 3}
	\label{eq:ScalingOptViscosity}
\end{equation}
for the optical shear viscosity. Repeating the scaling arguments as 
described in Section~\ref{sec:scaling} for two spatial dimensions, one time dimension and 
$1/2-\epsilon$ auxiliary time dimensions, the optical shear viscosity is 
expected to scale as
\begin{equation}
	\eta(\omega) \sim \omega^{(d + (1/2-\epsilon) z -\theta_\eta)/z},
\end{equation}
where $\theta_\eta$ is a hyperscaling violation exponent. The result in 
Eq.~\eqref{eq:ScalingOptViscosity} corresponds to $\theta_\eta = 3$, 
and thus $\theta_\eta \neq \theta$. The origin of this breakdown of the scaling 
expectation
is the $k_y^2$ factor in Eq.~\eqref{eq:TxyTxy}, which is dominated by 
contributions 
near the cutoff.

Instead, the result in Eq.~(\ref{eq:ScalingOptViscosity}) suggests
that the viscosity scales like a conductivity. For the conductivity, 
the arguments in Section~\ref{sec:scaling} imply that for the present dimensionally extended system,
the scaling law in Eq.~(\ref{sigmaQtheta}) is modified to
\begin{equation}
\sigma (\omega) \sim \omega^{(d + (1/2 - \epsilon)z - \theta - 2)/z};
\label{eq:sigmaomega}
\end{equation}
using the values $d=2$, $\theta=1$, and $z=3/(3 - 2 \epsilon)$, this agrees with
Eq.~(\ref{eq:ScalingOptViscosity}). The Ward identity analysis in Appendix~\ref{app:ward} shows that the 
identity of the scaling between the viscosity and the conductivity holds to all orders.

In the above computation, we considered the contributions to the optical 
viscosity from two patches on the Fermi surface. In Appendix~\ref{app:fullFermi}, we show 
that the scaling is the same if contributions from the full Fermi surface are 
taken into account. Moreover, by using Ward identities we trace the above 
conclusion back to the emergent rotation invariance of the low energy field 
theory, or equivalently to the fact that the Fermi surface curvature does not 
flow. 

Given the scaling of entropy in the present system
\begin{equation}
s \sim T^{(d + (1/2 - \epsilon)z - \theta)/z},
\end{equation}
our main result in Eq.~(\ref{etaos2}) would follow from Eq.~(\ref{eq:sigmaomega}) provided the viscosity
scaled in the same manner with $T$ in the regime $\omega \ll T$, as it does with $\omega$ in $\omega \gg T$.
We will turn to this important question in the following section.

\section{Boltzmann equation and DC viscosity}
\label{sec:dcviscosity}

This section presents a Boltzmann equation analysis which shows that $\omega/T$
scaling applies, and that the $\omega$-dependent results above can be extended 
to the d.c.~viscosity with $\omega \rightarrow T$. We set $N=1$ in this section for 
convenience. The DC viscosity may be derived in linear response by applying a static source 
that couples linearly to $T_{xy}$, which is equivalent to applying a static 
source that couples linearly to $J_y$ for the fermion contribution, i.e. a 
chiral electric field.

\begin{figure}[h]
	\centering
	\includegraphics[height=3.0in]{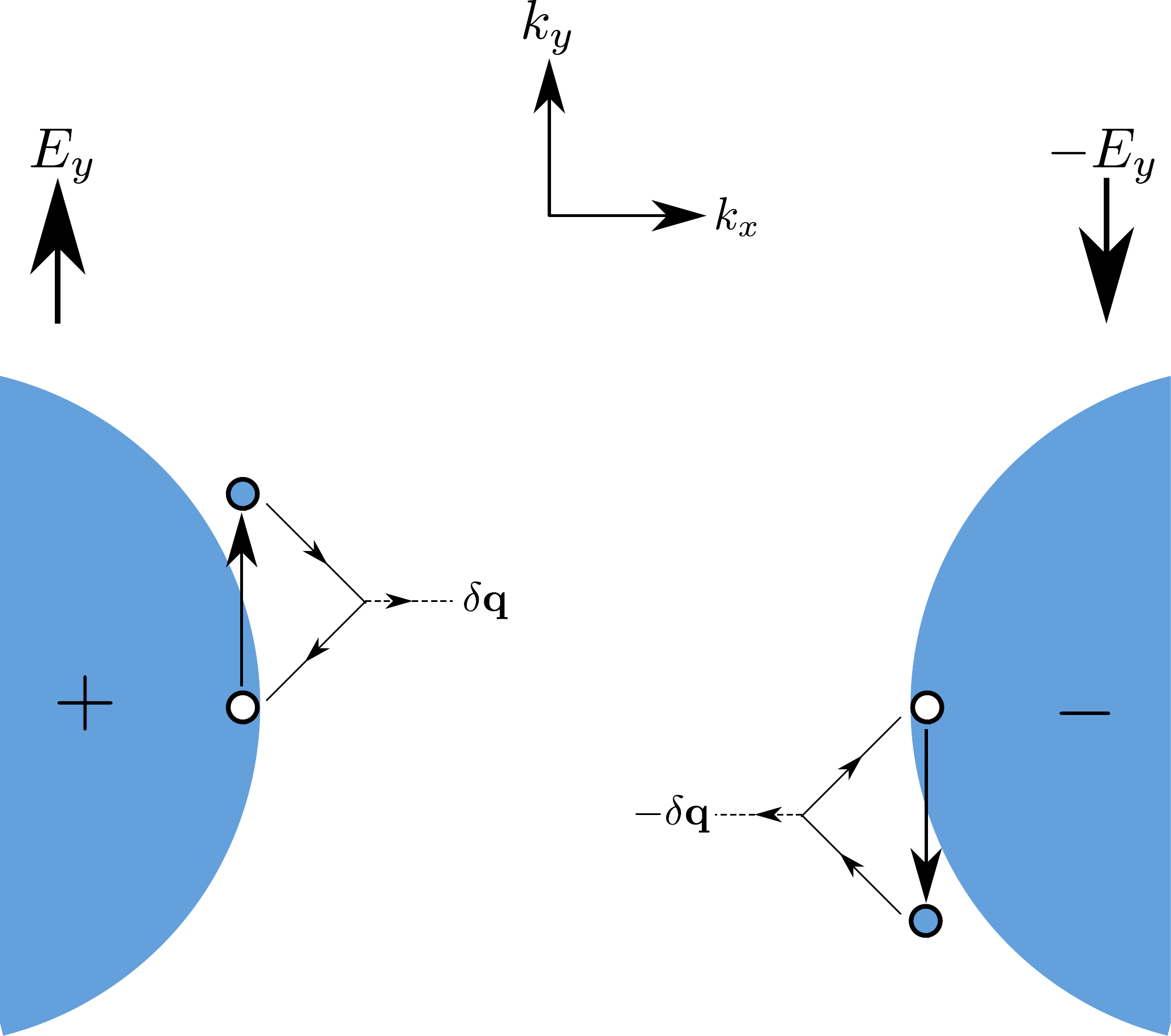}
	\caption{Elementary excitations due to the chiral electric field that carry a net chiral current at zero total momentum relative to the filled band in a two-patch system. The chiral current can decay via the emission of bosons of opposite momenta on the two patches. Since the individual bosons carry nonzero momentum, the boson distribution responds to the applied chiral electric field and is no longer in equilibrium unlike in a particle-hole symmetric system like those studied in Refs.~\onlinecite{fritz11, Patel2015}, where the bosons required to relax the elementary excitations have zero momentum.}
	\label{fig:Elprocess}
\end{figure}

Since our action is invariant under inversion for the full Fermi surface, i.e. 
\begin{equation}
\tilde{\psi}_s(k_0, k_x, k_y) = \tilde{\psi}_{\bar{s}}(k_0, -k_x, -k_y), 
\end{equation} 
and this leaves $J_y$ invariant but inverts the total momentum $P_i\rightarrow-P_i$, the chiral current has zero overlap with the conserved total momentum, i.e.
\begin{equation}
\chi_{J_yP_i} \equiv \int_0^{1/T} \langle J_y(\tau) P_i(0)\rangle = 0.
\end{equation}
Thus, the DC chiral conductivities and hence the DC viscosities are finite and can be determined using the Boltzmann equation. Fig.~\ref{fig:Elprocess} illustrates how chiral currents can be excited without changing the total momentum of the system. This requires oppositely directed electric fields to be applied to the two patches. 

The kinetic part of the fermion Hamiltonian in the dimensionally regularized theory may be diagonalized as
\begin{eqnarray}
&& H_f^0 =  \int \frac{d^d\mathbf{k}}{(2\pi)^d} \bar{\psi}(\mathbf{k}) \left[ i\bar{\mathbf{\Gamma}} \cdot \bar{\mathbf{K}} + i \gamma_{x} \delta_k \right] \psi(\mathbf{k})\nonumber \\ 
&&= \sum_{m=\pm} \int \frac{d^d\mathbf{k}}{(2\pi)^d} m \lambda^{\dagger}_{m}(\mathbf{k})\xi(\mathbf{k}) \lambda_{m}(\mathbf{k}),
\end{eqnarray}
where we use $\mathbf{k} = (k_x, k_y, \bar{\mathbf{K}})$, $\bar{\mathbf{K}}= (k_1,..., k_{d-2})$ and $\bar{\mathbf{\Gamma}} = (\gamma_1,..., \gamma_{d-2})$,
with the dispersion
\begin{equation}
\xi(\mathbf{k})=\left(\bar{\mathbf{K}}^2+\delta_k^2 \right)^{1/2}.
\end{equation}
The $y$-component of the chiral current density becomes
\begin{equation}
J_y= \left(\sum_{m=\pm} \int \frac{d^d\mathbf{k}}{(2\pi)^d} \frac{m \delta_k \partial_{k_y} \delta_k}{\sqrt{\bar{\mathbf{K}}^2+\delta_k^2}} \lambda^{\dagger}_m(\mathbf{k}) \lambda_m(\mathbf{k})\right)+J_y^{II},
\end{equation}
where $J_y^{II}$ contains particle-hole terms $\lambda^{\dagger}_+\lambda_-,~\lambda^{\dagger}_-\lambda_+$ that are unimportant for transport in the DC  regime of interest~\cite{FSMS08,fritz11}. Defining the non-equilibrium on-shell fermion distribution functions
\begin{equation}
f_f^m(t,\mathbf{k})=\langle\lambda^{\dagger}_{m}(t,\mathbf{k}) \lambda_{m}(t,\mathbf{k})\rangle,
\end{equation}
and the non-equilibrium off-shell boson distribution function $f_b(t,\mathbf{q},\Omega)$, we can write down the following collision equations in presence of an applied chiral electric field $\mathbf{E}$~\cite{Kamenev2011, fritz11, Patel2015}:
\begin{widetext}
\begin{align}
&\left(\frac{\partial}{\partial t} + \mathbf{E}\cdot\frac{\partial}{\partial\mathbf{p}}\right)f_f^m(\mathbf{p},t)= - e^2\mu^\epsilon\sum_{m^\prime=\pm}\int\frac{d^d\mathbf{q}}{(2\pi)^d}M_{mm^\prime}(\mathbf{p},\mathbf{q})\mathrm{Im}\left[D^R(\mathbf{p}-\mathbf{q},m\xi(\mathbf{p})-m^\prime\xi(\mathbf{q}))\right]\nonumber \\
&\times\Big\{f_f^m(t,\mathbf{p})(1-f_f^{m^\prime}(t,\mathbf{q}))+f_b(t,\mathbf{p}-\mathbf{q},m\xi(\mathbf{p})-m^\prime\xi(\mathbf{q}))(f_f^m(t,\mathbf{p})-f_f^{m^\prime}(t,\mathbf{q}))\Big\},  \\  \nonumber \\
&\left[\frac{\partial}{\partial\Omega}(2\Omega^2-\mathrm{Re}[\Sigma_b^R(t,\mathbf{q},\Omega)])\frac{\partial}{\partial t}+\frac{\partial\mathrm{Re}[\Sigma_b^R(t,\mathbf{q},\Omega)]}{\partial t}\frac{\partial}{\partial \Omega}\right]f_b(t,\mathbf{q},\Omega) = \nonumber \\
&4\pi e^2\mu^\epsilon\sum_{m,m^\prime=\pm}\int\frac{d^d\mathbf{k}}{(2\pi)^d}M_{mm^\prime}(\mathbf{k}+\mathbf{q},\mathbf{k})\delta(m\xi(\mathbf{k}+\mathbf{q})-m^\prime\xi(\mathbf{k})-\Omega)\Bigg[f_f^m(t,\mathbf{k}+\mathbf{q})(1-f_f^{m^\prime}(t,\mathbf{k})) \nonumber \\
&+f_b(t,\mathbf{q},\Omega)(f_f^m(t,\mathbf{k}+\mathbf{q})-f_f^{m^\prime}(t,\mathbf{k}))\Bigg],
\label{eq:ceqs}
\end{align}
where the interaction matrix elements are
\begin{equation}
M_{mm^\prime}(\mathbf{p},\mathbf{q}) = \frac{1}{2}\left(1 + mm^\prime\frac{\delta_p\delta_q-\bar{\mathbf{P}}\cdot\bar{\mathbf{Q}}}{\xi(\mathbf{p})\xi(\mathbf{q})}\right)
\end{equation}
(Note that $M_{++}=M_{--}$, $M_{+-}=M_{-+}$ and $M_{mm^\prime}(\mathbf{p},\mathbf{q}) = M_{mm^\prime}(\mathbf{q},\mathbf{p})$), and
\begin{equation}
D^R(\mathbf{k},\omega)=\frac{|k_y|}{|k_y|^3 + \beta_d e^2 \mu^\epsilon (\bar{\mathbf{K}}^2-\omega^2)^{(d-1)/2}},
\end{equation}
where $\beta_d$ depends only on $d$ and is free of poles in $\epsilon$~\cite{Dalidovich2013}. The additional self-energy component appearing in the boson collision equation is given by~\cite{Patel2015}
\begin{equation}
\mathrm{Re}[\Sigma_b^R(t,\mathbf{q},\Omega)] = -2e^2\mu^\epsilon\sum_{mm^\prime=\pm} \int \frac{d^d\mathbf{k}}{(2\pi)^d}M_{mm^\prime}(\mathbf{k},\mathbf{k}+\mathbf{q})\frac{f_f^{m^\prime}(t,\mathbf{k}+\mathbf{q})-f_f^m(t,\mathbf{k})}{m\xi(\mathbf{k})-m^\prime\xi(\mathbf{k}+\mathbf{q})+\Omega}.
\end{equation}
Both collision integrals vanish regardless of what $D^R$ is when the equilibrium distributions are used due to the identity
\begin{equation}
n_f(x)(1-n_f(y))+n_b(x-y)(n_f(x)-n_f(y)) = 0.
\label{eq:distidentity}
\end{equation}

We parameterize the deviations of the distributions from equilibrium in frequency space
\begin{align}
&f_f^m(\omega,\mathbf{p}) =  2\pi\delta(\omega)n_f(m\xi(\mathbf{p})) - Tn_f^\prime(\xi(\mathbf{p}))\varphi^m(\omega,\delta_p,\bar{\mathbf{P}})\mathbf{E}(\omega)\cdot\mathbf{\nabla_p}\xi(\mathbf{p}), \nonumber \\
&f_b(\omega,\mathbf{q}, \Omega) = 2\pi\delta(\omega)n_b(\Omega) + u(\omega,\mathbf{q},\Omega)|\mathbf{E}(\omega)|.
\end{align}
Using these, we linearize the collision equations with $\mathbf{E}=E_y\hat{e}_y$ as we are interested in $J_y$. In the DC limit, we obtain (since $\lim_{\omega\rightarrow0}(-i\omega+0^+)\varphi^m(\omega,\delta_p,\bar{\mathbf{P}})$ and $\lim_{\omega\rightarrow0}(-i\omega+0^+)u(\omega, \mathbf{q}, \Omega)$ are expected to vanish in the presence of interactions)
\begin{align}
&\frac{2m\delta_p\sqrt{d-1} p_y}{\xi(\mathbf{p})} n_f^\prime(\xi(\mathbf{p})) = -e^2\mu^\epsilon\sum_{m^\prime=\pm}\int\frac{d^d\mathbf{q}}{(2\pi)^d}M_{mm^\prime}(\mathbf{p},\mathbf{q})\mathrm{Im}\left[D^R(\mathbf{p}-\mathbf{q},m\xi(\mathbf{p})-m^\prime\xi(\mathbf{q}))\right] \nonumber \\
&\times\Big\{\frac{2\delta_q\sqrt{d-1} q_y}{\xi(\mathbf{q})}\varphi^{m^\prime}(\delta_q,\bar{\mathbf{Q}})Tn_f^\prime(\xi(\mathbf{q}))n_f(m\xi(\mathbf{p}))- \frac{2\delta_p\sqrt{d-1} p_y}{\xi(\mathbf{p})}\varphi^m(\delta_p,\bar{\mathbf{P}})Tn_f^\prime(\xi(\mathbf{p}))(1-n_f(m^\prime\xi(\mathbf{q}))) \nonumber \\
&+n_b(m\xi(\mathbf{p})-m^\prime\xi(\mathbf{q}))\left(\frac{2\delta_q\sqrt{d-1} q_y}{\xi(\mathbf{q})}\varphi^{m^\prime}(\delta_q,\bar{\mathbf{Q}})Tn_f^\prime(\xi(\mathbf{q}))-\frac{2\delta_p\sqrt{d-1} p_y}{\xi(\mathbf{p})}\varphi^m(\delta_p,\bar{\mathbf{P}})Tn_f^\prime(\xi(\mathbf{p}))\right) \nonumber \\
&+\mathrm{sgn}(E_y)u(\mathbf{p}-\mathbf{q},m\xi(\mathbf{p})-m^\prime\xi(\mathbf{q}))(n_f(m\xi(\mathbf{p}))-n_f(m^\prime\xi(\mathbf{q})))\Big\}.
\label{eq:fcel}
\end{align}
where we have suppressed the now zero frequency argument on the $\varphi$'s and $u$'s. For the boson collision equation we obtain
\begin{align}
&u(\mathbf{q},\Omega) = \mathrm{sgn}(E_y) \frac{I_1[\varphi,\mathbf{q},\Omega]}{I_2(\mathbf{q},\Omega)}, \nonumber \\
&I_1[\varphi,\mathbf{q},\Omega] =  4\pi e^2\mu^\epsilon\sum_{mm^\prime=\pm}\int\frac{d^d\mathbf{k}}{(2\pi)^d}M_{mm^\prime}(\mathbf{k}+\mathbf{q},\mathbf{k})\delta(m\xi(\mathbf{k}+\mathbf{q})-m^\prime\xi(\mathbf{k})-\Omega) \nonumber \\
&\times \Bigg[\Big\{\frac{2\delta_k\sqrt{d-1} k_y}{\xi(\mathbf{k})} \varphi^{m^\prime}(\delta_k, \bar{\mathbf{K}})Tn_f^\prime(\xi(\mathbf{k}))n_f(m\xi(\mathbf{k}+\mathbf{q})) \nonumber \\
&-\frac{2\delta_{k+q}\sqrt{d-1}(k_y+q_y)}{\xi(\mathbf{k}+\mathbf{q})}\varphi^m(\delta_{k+q},\bar{\mathbf{K}}+\bar{\mathbf{Q}}) T n_f^\prime(\xi(\mathbf{k}+\mathbf{q}))(1-n_f(m^\prime\xi(\mathbf{k})))\Big\}  \nonumber \\
&+n_b(\Omega)\Big\{\frac{2\delta_k\sqrt{d-1} k_y}{\xi(\mathbf{k})} \varphi^{m^\prime}(\delta_k, \bar{\mathbf{K}})Tn_f^\prime(\xi(\mathbf{k}))-2\frac{\delta_{k+q}\sqrt{d-1}(k_y+q_y)}{\xi(\mathbf{k}+\mathbf{q})}\varphi^m(\delta_{k+q},\bar{\mathbf{K}}+\bar{\mathbf{Q}}) T n_f^\prime(\xi(\mathbf{k}+\mathbf{q}))\Big\}\Bigg], \nonumber \\
&I_2(\mathbf{q},\Omega) = -4\pi e^2\mu^\epsilon\sum_{mm^\prime=\pm}\int\frac{d^d\mathbf{k}}{(2\pi)^d}M_{mm^\prime}(\mathbf{k}+\mathbf{q},\mathbf{k})\delta(m\xi(\mathbf{k}+\mathbf{q})-m^\prime\xi(\mathbf{k})-\Omega) \nonumber \\
&\times \Big\{n_f(m\xi(\mathbf{k}+\mathbf{q}))-n_f(m^\prime\xi(\mathbf{k}))\Big\}.
\label{eq:bcel}
\end{align}

Since the driving term for the fermions in Eq.~(\ref{eq:ceqs}) is of opposite signs for the $+$ and $-$ quasiparticles, we expect $\varphi^m(\delta_{p},\bar{\mathbf{P}}) = m\varphi(\delta_{p},\bar{\mathbf{P}})$. Then, using the properties of the matrix elements $M$ noted previously and that $n_{f,b}(x)+n_{f,b}(-x) = \pm 1$, one can see that $u$ is an odd function of $\Omega$ and hence that the same $\varphi(\delta_{k},\bar{\mathbf{K}})$ can be used to solve the collision equations for both branches of quasiparticles.

In the (convergent) boson collision integrals in Eq.~(\ref{eq:bcel}), we shift $k_x\rightarrow k_x - \sqrt{d-1}k_y^2$ and integrate over $k_y$. In the (also convergent) fermion collision integral Eq.~(\ref{eq:fcel}), after inserting $u$ derived from the boson collision equation we shift $q_x\rightarrow q_x - \sqrt{d-1}q_y^2$ followed by $q_y\rightarrow q_y+p_y$, and then integrate out $q_y$ after dividing through by $2\sqrt{d-1}p_y$. Terms that are odd in $q_y$ drop out, and we are left with
\begin{align}
&\frac{m\delta_p}{\xi(\mathbf{p})} n_f^\prime(\xi(\mathbf{p})) = -\frac{e^2\mu^\epsilon}{2}\sum_{m^\prime=\pm}\int\frac{d^{d-1}\mathbf{q}}{(2\pi)^d}\left(1 + mm^\prime\frac{\delta_p q_x-\bar{\mathbf{P}}\cdot\bar{\mathbf{Q}}}{\xi(\mathbf{p})\sqrt{\bar{\mathbf{Q}}^2+q_x^2}}\right) \nonumber \\
&\times\mathrm{Im}\left[\frac{4\pi}{\sqrt{27}}\left(\beta_d e^2\mu^\epsilon\left((\bar{\mathbf{P}}-\bar{\mathbf{Q}})^2-\left(m\xi(\mathbf{p})-m^\prime\sqrt{\bar{\mathbf{Q}}^2+q_x^2}\right)^2\right)^{(d-1)/2}\right)^{-1/3}\right] \nonumber \\
&\times\Bigg\{\frac{q_x}{\sqrt{\bar{\mathbf{Q}}^2+q_x^2}}\varphi^{m^\prime}(q_x,\bar{\mathbf{Q}})Tn_f^\prime\left(\sqrt{\bar{\mathbf{Q}}^2+q_x^2}\right)n_f(m\xi(\mathbf{p})) \nonumber \\
&- \frac{\delta_p}{\xi(\mathbf{p})}\varphi^m(\delta_p,\bar{\mathbf{P}})Tn_f^\prime(\xi(\mathbf{p}))\left(1-n_f\left(m^\prime\sqrt{\bar{\mathbf{Q}}^2+q_x^2}\right)\right) \nonumber \\
&+n_b\left(m\xi(\mathbf{p})-m^\prime\sqrt{\bar{\mathbf{Q}}^2+q_x^2}\right)\left(\frac{q_x}{\sqrt{\bar{\mathbf{Q}}^2+q_x^2}}\varphi^{m^\prime}(q_x,\bar{\mathbf{Q}})Tn_f^\prime\left(\sqrt{\bar{\mathbf{Q}}^2+q_x^2}\right)-\frac{\delta_p}{\xi(\mathbf{p})}\varphi^m(\delta_p,\bar{\mathbf{P}})Tn_f^\prime(\xi(\mathbf{p}))\right) \nonumber \\
&-\frac{H_1[\varphi,\mathbf{\bar{P}}-\mathbf{\bar{Q}},m\xi(\mathbf{p})-m^\prime\sqrt{\bar{\mathbf{Q}}^2+q_x^2}]}{H_2(\mathbf{\bar{P}}-\mathbf{\bar{Q}},|m\xi(\mathbf{p})-m^\prime\sqrt{\bar{\mathbf{Q}}^2+q_x^2}|)}\left(n_f(m\xi(\mathbf{p}))-n_f\left(m^\prime\sqrt{\bar{\mathbf{Q}}^2+q_x^2}\right)\right)\Bigg\},
\label{eq:fcefin}
\end{align}
where 
\begin{align}
&H_1[\varphi,\bar{\mathbf{Q}},\Omega] = \sum_{mm^\prime s}\int\frac{d^{d-1}\mathbf{k}}{(2\pi)^d}\frac{|m^\prime\sqrt{\bar{\mathbf{K}}^2+k_x^2}+\Omega|\Theta((m^\prime\sqrt{\bar{\mathbf{K}}^2+k_x^2}+\Omega)^2-(\bar{\mathbf{K}}+\bar{\mathbf{Q}})^2)}{\left((m^\prime\sqrt{\bar{\mathbf{K}}^2+k_x^2}+\Omega)^2-(\bar{\mathbf{K}}+\bar{\mathbf{Q}})^2\right)^{1/2}} \nonumber \\
&\times \left(1 + mm^\prime\frac{s\left((m^\prime\sqrt{\bar{\mathbf{K}}^2+k_x^2}+\Omega)^2-(\bar{\mathbf{K}}+\bar{\mathbf{Q}})^2\right)^{1/2} k_x-(\bar{\mathbf{K}}+\bar{\mathbf{Q}})\cdot\bar{\mathbf{K}}}{|m^\prime\sqrt{\bar{\mathbf{K}}^2+k_x^2}+\Omega|\sqrt{\bar{\mathbf{K}}^2+k_x^2}}\right) \nonumber \\
&\times\Bigg[\varphi^m(s((m^\prime(\bar{\mathbf{K}}^2+k_x^2)^{1/2}+\Omega)^2-(\bar{\mathbf{K}}+\bar{\mathbf{Q}})^2)^{1/2},\bar{\mathbf{K}}+\bar{\mathbf{Q}})\frac{s((m^\prime(\bar{\mathbf{K}}^2+k_x^2)^{1/2}+\Omega)^2-(\bar{\mathbf{K}}+\bar{\mathbf{Q}})^2)^{1/2}}{|m^\prime\sqrt{\bar{\mathbf{K}}^2+k_x^2}+\Omega|} \nonumber \\
&\times Tn_f^\prime(m^\prime(\bar{\mathbf{K}}^2+k_x^2)^{1/2}+\Omega)\Big\{1-n_f(m^\prime(\bar{\mathbf{K}}^2+k_x^2)^{1/2})+n_b(\Omega)\Big\} \nonumber \\
&-\varphi^{m^\prime}(k_x,\bar{\mathbf{K}})\frac{k_x}{\sqrt{\bar{\mathbf{K}}^2+k_x^2}}Tn_f^\prime(m^\prime(\bar{\mathbf{K}}^2+k_x^2)^{1/2})\Big\{n_f(m^\prime(\bar{\mathbf{K}}^2+k_x^2)^{1/2}+\Omega)+n_b(\Omega)\Big\}\Bigg], \nonumber \\
\nonumber \\
&H_2(\bar{\mathbf{Q}},\Omega) = \sum_{mm^\prime s}\int\frac{d^{d-1}\mathbf{k}}{(2\pi)^d}\frac{|m^\prime\sqrt{\bar{\mathbf{K}}^2+k_x^2}+\Omega|\Theta((m^\prime\sqrt{\bar{\mathbf{K}}^2+k_x^2}+\Omega)^2-(\bar{\mathbf{K}}+\bar{\mathbf{Q}})^2)}{\left((m^\prime\sqrt{\bar{\mathbf{K}}^2+k_x^2}+\Omega)^2-(\bar{\mathbf{K}}+\bar{\mathbf{Q}})^2\right)^{1/2}} \nonumber \\
&\times \left(1 + mm^\prime\frac{s\left((m^\prime\sqrt{\bar{\mathbf{K}}^2+k_x^2}+\Omega)^2-(\bar{\mathbf{K}}+\bar{\mathbf{Q}})^2\right)^{1/2} k_x-(\bar{\mathbf{K}}+\bar{\mathbf{Q}})\cdot\bar{\mathbf{K}}}{|m^\prime\sqrt{\bar{\mathbf{K}}^2+k_x^2}+\Omega|\sqrt{\bar{\mathbf{K}}^2+k_x^2}}\right) \nonumber \\
&\times\left\{n_f\left(m^\prime\sqrt{\bar{\mathbf{K}}^2+k_x^2}+\Omega\right)-n_f\left(m^\prime\sqrt{\bar{\mathbf{K}}^2+k_x^2}\right)\right\}.
\label{eq:fcefinreps}
\end{align}

If we choose $\varphi^m(\delta_p,\bar{\mathbf{P}})= 
\varphi(\delta_p,\bar{\mathbf{P}})$ with $\varphi(\delta_p,\bar{\mathbf{P}}) = 
C(T)(\delta_p^2+\bar{\mathbf{P}}^2)^{1/2}/\delta_p$, the right hand side of 
Eq.~(\ref{eq:fcefin}) vanishes due to the identity
\begin{equation}
n_f^\prime(x)(n_f(y)-1)+n_f(x)n_f^\prime(y)-n_b(x-y)(n_f^\prime(x)-n_f^\prime(y)) = 0.
\label{eq:distderidentity}
\end{equation}
This is the zero mode of the collision equation, and will lead to an infinite conductivity if excited. However, this mode cannot be excited by the \textit{chiral} electric field as it produces the \textit{same} (instead of \textit{opposite}) deviation in the $+$ and $-$ quasiparticle distributions. This mode will be excited by a \textit{normal} electric field, and is responsible for the infinite DC \textit{charge} conductivity of the system. The modes excited by the chiral electric field obey $\varphi^m(\delta_{p},\bar{\mathbf{P}}) = m\varphi(\delta_{p},\bar{\mathbf{P}})$ and are orthogonal to the zero mode, yielding a finite \textit{chiral} conductivity (or viscosity).

We have
\begin{equation}
\eta \sim \sigma_{yy} =\frac{J_y}{E_y} = 8(1-d)T \int dp_y p_y^2 \int \frac{d^{d-1}\mathbf{p}}{(2\pi)^d}\frac{p_x^2}{\bar{\mathbf{P}}^2+p_x^2}n_f^\prime((\bar{\mathbf{P}}^2+p_x^2)^{1/2}) \varphi(p_x,\bar{\mathbf{P}}),
\label{eq:eta}
\end{equation}
\end{widetext}

where we shifted $p_x\rightarrow p_x-\sqrt{d-1}p_y^2$. Counting powers in Eq.~(\ref{eq:fcefin}), we obtain
\begin{equation}
\varphi(\xi(\mathbf{p}),\bar{\mathbf{P}}) = \beta_d^{1/3}e^{-4/3}\mu^{-2\epsilon/3}T^{-2(d-1)/3-1}\tilde{\varphi}(\xi(\mathbf{p})/T, \bar{\mathbf{P}}/T).
\end{equation} 
Inserting this into Eq.~(\ref{eq:eta}) and using the fixed point values of $z^\ast = 3/(2(d-1))$ and $e^{\ast 4/3}\propto \epsilon$~\cite{Dalidovich2013}, we have, to leading order in $\epsilon$,
\begin{equation}
\eta\sim \frac{1}{\epsilon} T^{-1/z + d-2} \int dp_y p_y^2, 
\end{equation}
which is the expected quantum critical scaling.

If the Boltzmann analysis at this order is performed directly in $d=2$, then the 
collision equations are solved exactly by using the collisionless 
momentum-independent solution for $\varphi$, and thus collisions with the boson 
do not induce a finite DC viscosity. The reason for this is purely kinematic, 
stemming from the special structure of the patch dispersions in $d=2$ which have 
Galilean invariance in the $y$ direction and a constant $x$ velocity and was 
noted earlier in Ref.~\onlinecite{Maslov2011}. The quantum critical scaling could 
possibly be restored by appropriately resumming contributions at higher orders 
in perturbation theory.

\section{Conclusions}
\label{sec:conc}

This paper has exposed the unconventional scaling of the shear viscosity in a theory with a critical Fermi surface. For the Ising-nematic QCP in $d=2$, we computed the optical and DC viscosities in an expansion in $\epsilon = 5/2 - d$ below the upper critical dimension, and showed that the viscosity scales differently than expected from that of a critical point with an effective reduced dimensionality of $(d-1)$-dimensional excitations transverse to the Fermi surface. As a consequence, the ratio $\eta/s$ diverges at low temperatures as $T^{-2/z}$ instead of saturating the universal bound like in other strongly-coupled field theories in the literature. We expect that this is a general phenomenon of metallic quantum critical states where hyperscaling is violated due to the presence of a critical Fermi surface, including states described by Fermi surfaces coupled to gauge fields. 
However, we do expect
that metallic critical points associated with singular `hot spots' on the Fermi surface \cite{Patel2015} will have a finite $\eta/s$, up to logarithmic factors.

\section*{Acknowledgements}
We thank R. Davison and W. Witczak-Krempa for valuable discussions.
This research was supported by the NSF under Grant DMR-1360789 and MURI grant W911NF-14-1-0003 from ARO.
Research at Perimeter Institute is supported by the Government of Canada through Industry Canada and by the Province of Ontario 
through the Ministry of Research and Innovation. AE acknowledges support from 
the German National Academy of Sciences Leopoldina through grant LPDS~2014-13. 
SS also acknowledges support from Cenovus Energy at Perimeter Institute.
AE and AAP contributed equally to this work.

\appendix

\begin{widetext}
\section{Optical viscosity: two-loop computations}
\label{app:twoloop}

The two-loop self-energy correction to the stress tensor autocorrelation 
function is given by
\begin{equation}
\begin{split}
	\langle T_{xy} T_{xy}\rangle_\text{SE} (i\omega) &= 2 N \intd{k}{d+1} k_y^2  
\operatorname{tr} \Bigl(\gamma_x G_0(k+q) \gamma_x G_0(k) 
\Sigma_1(k) G_0(k)\Bigr)\\
	&= 4 e^{4/3} u_{\Sigma,0} \epsilon^{-1} \intd{k}{d+1} k_y^2 
\Bigl(\frac{\mu}{|\bs K|}\Bigr)^{2\epsilon/3} \frac{2 \delta_k^2 \bs K^2 + \bs K 
\cdot (\bs K + \bs Q) (\delta_k^2 - \bs K^2)}{\big((\bs K + \bs Q)^2 + 
\delta_k^2)(\bs K^2 + \delta_k^2)^2},
\end{split}
\end{equation}
where we only kept the pole contribution to the self-energy and set $\epsilon 
= 0$ in the prefactor $u_{\Sigma,\epsilon = 0} = (2 \cdot 6^{1/3} \pi)^{-1}$. 
The self-energy correction can be computed using Feynman parameters. The 
integral is first rewritten as
\begin{equation}
	\langle T_{xy} T_{xy}\rangle_\text{SE}(i\omega) =4 
(e^2\mu^\epsilon)^\frac{2}{3} u_{\Sigma,0} \epsilon^{-1} \intd{k}{d+1} k_y^2
\intdefnopi{x}{0}{1} \frac{1-x}{|\bs K|^\frac{2\epsilon}{3}} \frac{2\delta_k^2 
\bs K^2 + \bs K \cdot (\bs K + \bs Q)(\delta_k^2 - \bs K^2)}{\bigl[x (\bs K + 
\bs Q)^2 + (1-x) \bs K^2 + \delta_k^2\bigr]^3}.
\end{equation}
Eliminating $k_y$ from the fraction by a variable shift of $k_x$ and subsequent 
integration over $k_x$ yield
\begin{equation}
\begin{split}
	=\frac{\Gamma(3)}{4}(e^2\mu^\epsilon)^\frac{2}{3} u_{\Sigma,0} \epsilon^{-1} 
&\intdef{k_y}{}{} k_y^2 \intd{K}{d-1} \intdefnopi{x}{0}{1} \frac{1-x}{|\bs 
K|^\frac{2\epsilon}{3}} \times \Bigl[\frac{3\bs K^2 + \bs K \cdot \bs 
Q}{\bigl[\bs 
K^2 + x (2\bs K \cdot \bs Q + \bs Q^2)\bigr]^\frac{3}{2}} - \frac{3 \bs K^2 
(\bs K^2 + \bs K \cdot \bs Q)}{\bigl[\bs K^2 + x (2 \bs K \cdot \bs Q + \bs 
Q^2)\bigr]^\frac{5}{2}}\Bigr].
\end{split}
\end{equation}
Again using Feynman parameters to rewrite the products in the integrand, we 
obtain
\begin{equation}
\begin{split}
	=\frac{\Gamma(3)}{4\Gamma(\frac{\epsilon}{3})}& (e^2\mu^\epsilon)^\frac{2}{3} 
u_{\Sigma,0} \epsilon^{-1} 
\intdef{k_y}{}{} k_y^2 \intd{K}{d-1}\intdefnopi{x}{0}{1}\intdefnopi{y}{0}{1} 
\Bigl[\frac{\Gamma(\frac{9+2\epsilon}{6})}{\Gamma(\frac{3}{2})} 
\frac{(1-x) y^{\frac{\epsilon}{3}-1} (1-y)^\frac{1}{2} (3 \bs K^2 + \bs K \cdot 
\bs Q)}{\bigl[\bs K^2 + x (1-y) (2\bs K \cdot \bs Q + \bs 
Q^2)]^{\frac{3}{2}+\frac{\epsilon}{3}}} \\
	&\quad - 
\frac{\Gamma(\frac{15+2\epsilon}{6})}{\Gamma(\frac{5}{2})} \frac{3 (1-x) 
y^{\frac{\epsilon}{3}-1} (1-y)^\frac{3}{2} \bs K^2 (\bs K^2 + \bs K \cdot \bs 
Q)}{\bigl[\bs K^2 + x (1-y) (2 \bs K \cdot \bs Q + \bs Q^2)\bigr]^{\frac{5}{2} 
+ \frac{\epsilon}{3}}}\Bigr].
\end{split}
\end{equation}
Completing squares in the denominator as
\begin{equation}
	\bs K^2 + x (1-y)(2 \bs K \cdot \bs Q + \bs Q^2) = (\bs K + x(1-y)\bs Q)^2 + 
x (1-y) (1-x + xy) \bs Q^2,
\end{equation}
 shifting $\bs K \rightarrow \bs K - x (1-y) \bs Q$, and neglecting terms that 
vanish due to symmetries when performing the $\bs K$-integration, we obtain
\begin{equation}
\begin{split}
	=&\frac{\Gamma(3)}{4\Gamma(\frac{\epsilon}{3})} (e^2 \mu^\epsilon)^\frac{2}{3} 
u_{\Sigma,0} \epsilon^{-1} \intdef{k_y}{}{} k_y^2 \intd{K}{d-1} 
\intdefnopi{x}{0}{1} \intdefnopi{y}{0}{1} (1-x) 	y^{\frac{\epsilon}{3}-1}\\
	& \times \Bigl\{\frac{\Gamma(\frac{9+2\epsilon}{6})}{\Gamma(\frac{3}{2})} 
(1-y)^\frac{1}{2} \frac{3 \bs K^2 - x (1-y) (1-3x(1-y)) \bs Q^2}{\bigl[\bs K^2 
+ 
x (1-y) (1-y+xy) \bs Q^2\bigr]^{\frac{3}{2}+\frac{\epsilon}{3}}} \\
	&\quad - \frac{\Gamma(\frac{15+2\epsilon}{6})}{\Gamma(\frac{5}{2})} 
\frac{3 (1-y)^\frac{3}{2}}{\bigl[\bs K^2 + 
x(1-y) (1-x+xy) \bs Q^2\bigr]^{\frac{5}{2} + \frac{\epsilon}{3}}} 
\Bigl[\bs K^4 - x(1-y)(1-2x(1-y)) \bs K^2 \bs Q^2 \\
	& \quad \quad \quad - 2 x (1-y) (1-2x 
(1-y)) (\bs K \cdot \bs Q)^2 - x^3 (1-y)^3 (1-x(1-y)) \bs Q^4\Bigr]\Bigr\}
\end{split}
\end{equation}
The remaining integrals can easily be computed using \verb|Mathematica|. First 
integrating over $\bs K$ and subsequently over $x$ and $y$, the pole 
contribution to the two-loop self-energy correction reads
\begin{equation}
	\langle T_{xy} T_{xy} \rangle_\text{SE}(i\omega) = e^{4/3} 
\epsilon^{-1} \intdef{k_y}{}{} k_y^2 |\omega|^{\frac{1}{2} - \epsilon} 
\Bigl(\frac{\mu}{|\omega|}\Bigr)^{2\epsilon/3} a_{\Sigma,0},
\end{equation}
where $a_{\Sigma,0} = u_{\text{1Loop},0} u_{\Sigma,0}$, after setting $\epsilon$ 
to zero in the numerical prefactors. 

\section{Relating conductivities and viscosities using Ward identities}
\label{app:ward}

The result in the main text, that the optical viscosity 
and optical conductivity scale in the same way, is not consistent with 
hyperscaling with an effectively reduced dimension. In 
order to substantiate this result, in the following we establish relations 
between the two transport quantities based on Ward identities.

The action for the patch theory of the Ising-nematic QCP in $d = 2$, 
Eq.~\eqref{eq:Action2D}, is invariant under an emergent rotational 
symmetry~\cite{Metlitski2010a},
\begin{gather}
	\Phi(q_0, q_x, q_y) \rightarrow \Phi'(q_0, q_x, q_y) = \Phi(q_0, q_x - \Theta 
q_y, q_y)\\
	\tilde\psi_{sj}(k_0, k_x, k_y) \rightarrow \tilde\psi_{sj}'(k_0, k_x, k_y) = 
\tilde\psi_{sj}\Bigl(k_0, k_x - \Theta k_y - s \frac{\Theta^2}{4}, k_y + s 
\frac{\Theta}{2}\Bigr).
\end{gather}
We will show that this symmetry restricts the 
scaling behavior of transport properties as a function of frequency. 
Starting from this transformation law, we derive a Ward identity for the 
generating functional of connected correlation functions~\cite{Negele1998},
\begin{gather}
	\mathcal G[\eta^\dagger, \eta, \phi] = \operatorname{ln} Z[\eta^\dagger, 
\eta, \phi]\\
	Z[\eta^\dagger, \eta, \phi] = \int D(\tilde \psi^\dagger, \tilde\psi) D(\Phi) 
e^{-S[\tilde\psi^\dagger, \tilde\psi, \Phi] - \int_k \sum_{s,j} 
(\tilde\psi^\dagger_{sj}(k) \eta_{sj}(k) + \eta^\dagger_{sj}(k) 
\tilde\psi_{sj}(k)) - \int_q \Phi(-q) \phi(q)},
\end{gather}
where $\eta^{(\dagger)}$ and $\phi$ are Grassmann and real source fields, 
respectively. Invariance under the above rotational symmetry implies
\begin{equation}
	\mathcal G[\eta'^\dagger, \eta', \phi'] = \mathcal G[\eta^\dagger, \eta, 
\phi],
\end{equation}
where the source fields transform as the physical fields. Differentiation with 
respect to $\Theta$ yields
\begin{equation}
	\frac{d}{d\Theta}\mathcal G[\eta'^\dagger, \eta', \phi'] = 0,
\end{equation}
which leads to the functional Ward identity
\begin{equation}
\begin{split}
	\intd{k}{3} \sum_{s,j} \Bigl\{&\Bigl[\bigl(k_y \partial_{k_x} 
\eta^\dagger_{sj}(k) 
- \frac{s}{2} \partial_{k_y} \eta^\dagger_{sj}(k)\bigr) \frac{\delta 
Z[\eta^\dagger, \eta, \phi]}{\delta \eta^\dagger_{sj}(k)} + \bigl(k_y 
\partial_{k_x} \eta_{sj}(k) - 
\frac{s}{2} \partial_{k_y} \eta_{sj}(k)\bigr) \frac{\delta 
Z[\eta^\dagger, \eta, \phi]}{\delta \eta_{sj}(k)}\Bigr]\\
	& - \int_q q_y \partial_{q_x}\phi(q) \frac{\delta 
Z[\eta^\dagger, \eta, \phi]}{\delta \phi(q)}\Bigr\} = 0.
\end{split}
\label{eq:FuncWI}
\end{equation}
In the following we are only interested in Ward identities for fermionic 
correlation functions and thus set $\phi = 0$ from the outset.

As an example how this functional Ward identity restricts correlation 
functions, we derive the Ward identity that follows from rotational symmetry 
for the fermionic Green's function. After computing suitable functional 
derivatives, we obtain
\begin{equation}
	\bigl(p_y \partial_{p_x} - \frac{s}{2} \partial_{p_y}\bigr) \frac{\delta^2 
Z[\eta^\dagger, \eta, 
0]}{\partial\eta_{sj}(p)\partial\eta^\dagger_{sj}(p)}\Bigl|_{\eta=\eta^\dagger 
= 0} = -\bigl(p_y \partial_{p_x} - \frac{s}{2}\partial_{p_y}\bigr) G_s(p) = 0,
\end{equation}
where $G_s(p)$ is the full fermionic Green's function. This is a partial 
differential equation for the momentum dependence of the latter. It can easily 
be verified that the Ward identity is fulfilled for $G_s(p) = G_s(p_0, s p_x + 
p_y^2)$, as expected.

For $q = q_0 \bs e_0 \neq 0$, the current-current correlation functions for the 
chiral current can be written as
\begin{gather}
\begin{split}
	\langle J_x(q) J_x(-q)\rangle &= \intd{k}{3}\intd{k'}{3} \sum_{j,s,j',s'} 
\langle \tilde\psi_{js}^\dagger(k+q) \tilde\psi_{js}(k) 
\tilde\psi_{j's'}^\dagger(k'-q) \tilde\psi_{j's'}(k')\rangle\\
	&= Z^{-1} \intd{k}{3} \intd{k'}{3} \sum_{j,s,j',s'} \frac{\delta^4 
Z[\eta^\dagger,\eta,0]}{\delta\eta_{js}(k+q)\delta\eta_{js}^\dagger(k) 
\delta\eta_{j's'}(k'-q) 
\delta\eta_{j's'}^\dagger(k')}\Bigr|_{\eta=\eta^\dagger=0}
\label{eq:JxJx}
\end{split}\\
\begin{split}
	\langle J_y(q) J_y(-q)\rangle &= 4\langle T_{xy}(q) T_{xy}(-q)\rangle\\
	&= 4\intd{k}{3}\intd{k'}{3} \sum_{j,s,j',s'} s s' k_y k_y'
\langle \tilde\psi_{js}^\dagger(k+q) \tilde\psi_{js}(k) 
\tilde\psi_{j's'}^\dagger(k'-q) \tilde\psi_{j's'}(k')\rangle\\
	&= 4 Z^{-1} \intd{k}{3} \intd{k'}{3} \sum_{j,s,j',s'} s s' k_y k_y' 
\frac{\delta^4 
Z[\eta^\dagger,\eta,0]}{\delta\eta_{js}(k+q)\delta\eta_{js}^\dagger(k) 
\delta\eta_{j's'}(k'-q) 
\delta\eta_{j's'}^\dagger(k')}\Bigr|_{\eta=\eta^\dagger=0}.
\label{eq:JyJy}
\end{split}
\end{gather}
Applying functional derivatives to the functional Ward identity 
Eq.~\eqref{eq:FuncWI}, we obtain a Ward identity for two-particle Green's 
functions,
\begin{equation}
	\bigl[(k_y \partial_{k_x} - \frac{s}{2} \partial_{k_y}) + (k_y' 
\partial_{k_x'} - \frac{s'}{2} \partial_{k_y'})\bigr]\frac{\delta^4 
Z[\eta^\dagger,\eta,0]}{\delta\eta_{js}(k+q)\delta\eta_{js}^\dagger(k) 
\delta\eta_{j's'}(k'-q) 
\delta\eta_{j's'}^\dagger(k')}\Bigr|_{\eta=\eta^\dagger=0} = 0.
\end{equation}
Using the method of characteristics, we can show that this Ward identity 
restricts the dependence of two-particle Green's function on spatial momenta as 
\begin{equation}
	\frac{\delta^4 
Z[\eta^\dagger,\eta,0]}{\delta\eta_{js}(k+q)\delta\eta_{js}^\dagger(k) 
\delta\eta_{j's'}(k'-q) 
\delta\eta_{j's'}^\dagger(k')}\Bigr|_{\eta=\eta^\dagger=0} = F_{js;j's'}(k_0, 
k_0', q_0; s k_x + k_y^2, s' k_x' + k_y'^2, s'k_y - s k_y'),
\end{equation}
analogously to the Ward identity for the one-particle Green's function. 

Inserting this result in Eqs.~\eqref{eq:JxJx} and~\eqref{eq:JyJy}, shifting and 
renaming integration variables, we obtain for the $J_x$ correlator
\begin{align}
	\langle J_x(q_0) J_x(-q_0) \rangle &= Z^{-1} \intd{k}{3} \intd{k'}{3} 
\sum_{j,s,j',s'} F_{js;j's'}(k_0, k_0', q_0; s k_x + k_y^2, s' k_x' + k_y'^2, 
s' k_y - s k_y') \nonumber\\
&= Z^{-1} \intd{k}{3} \intd{k'}{3} \sum_{j,s,j',s'} F_{js;j's'}(k_0, k_0', 
q_0; k_x, k_x', k_y).
\end{align}
Note that $k_y'$ does not appear in the integrand. For the $J_y$ correlator we 
obtain
\begin{align}
	\langle J_y(q_0) J_y(-q_0) \rangle &= 4 Z^{-1} \intd{k}{3}\intd{k'}{3} 
\sum_{j,s,j',s'} s s' k_y k_y' F_{js;j's'}(k_0, k_0', q_0; s k_x + k_y^2, s' 
k_x' + k_y'^2, s' k_y - s k_y') \nonumber\\
&= 4 Z^{-1} \intd{k}{3}\intd{k'}{3} \sum_{j,s,j',s'} k_y k_y' 
F_{js;j's'}(k_0, k_0', q_0; k_x, k_x', k_y - k_y')\nonumber\\
	&= 2 Z^{-1} \intd{k}{3}\intd{k'}{3} \sum_{j,s,j',s'} (k_y + k_y') k_y' 
F_{js;j's'}(k_0, k_0', q_0; k_x, k_x', k_y)\nonumber\\
	&\quad + 2 Z^{-1} \intd{k}{3}\intd{k'}{3} \sum_{j,s,j',s'} k_y (k_y + k_y') 
F_{js;j's'}(k_0, k_0', q_0; k_x, k_x', - k_y')
\end{align}
where in the last step we shifted $k_y \rightarrow k_y + k_y'$ and $k_y' 
\rightarrow k_y' + k_y$ in the first and second term, respectively. Replacing 
$k_y' \rightarrow -k_y'$ and subsequently renaming $k_y \leftrightarrow k_y'$ in 
the second term, the contributions $\sim k_y k_y'$ cancel and we hence obtain
\begin{equation}
 = 4 Z^{-1} \intd{k}{3}\intd{k'}{3} 
\sum_{j,s,j',s'} k_y'^2 F_{js;j's'}(k_0, k_0', q_0; k_x, k_x', k_y).
\end{equation}

Using the Ward identity for the emergent rotational symmetry of the patch 
theory, we have thus established that
\begin{align}
	\langle J_x(q_0) J_x(-q_0) \rangle = \frac{\intd{k_y}{}}{4 \intd{k_y}{} 
k_y^2} \langle J_y(q_0) J_y(-q_0) \rangle = \frac{\intd{k_y}{}}{\intd{k_y}{} 
k_y^2} \langle T_{xy}(q_0) T_{xy}(-q_0) \rangle.
\end{align}
As the Ward identity imposes restrictions only on the momentum dependence of 
the two-particle Green's function, this result is also valid in $d = 5/2 - 
\epsilon$. The above result implies that the optical viscosity and the optical 
(chiral) conductivity have the same frequency dependence,
\begin{equation}
	\sigma(\omega) \sim \eta(\omega).
\end{equation}
Using the results from Ref.~\onlinecite{Eberlein2016}, we obtain
\begin{equation}
 \sigma(\omega) \sim \eta(\omega) \sim \omega^{-1/2 - \epsilon / 3} 
\stackrel{\epsilon = 1/2}{=} \omega^{-2/3},
\label{eq:ScalingOptViscosity2}
\end{equation}
in agreement with the field theoretic result in Eq.~(\ref{eq:ScalingOptViscosity}).

\section{Contributions from the full Fermi surface}
\label{app:fullFermi}

The result Eq.~(\ref{eq:ScalingOptViscosity2}) for the scaling of the optical
viscosity was derived in the patch theory. The question arises whether the 
scaling could be different for the full Fermi surface, for example due to some 
preferred direction. In this section we show that this is not the case, using 
$d = 2$ for simplicity.

We first analyze how the $T_{xy}$ correlator transforms 
under the emergent rotation symmetry of the patch theory~\cite{Metlitski2010a}. 
In $d = 2$, $T_{xy}$ is given by
\begin{equation}
	T_{xy}(q) = \sum_{j, s, s'} \intm{k} \Bigl(k_y + \frac{q_y}{2}\Bigr) 
\tilde \psi_{sj}^\dagger(k+q) \sigma_{z,ss'} \tilde \psi_{s' j}(k).
\end{equation}
We obtain
\begin{align}
\langle T_{xy} T_{xy}\rangle_\text{1Loop}(q) \stackrel{q = \omega \bs e_0}{=} 
-N \intm{k} k_y^2 \operatorname{tr}\bigl(G_0(k_0 + \omega, \bs k) G_0(k_0, \bs 
k)\bigr)
\label{eq:TxyTxy_1loop_Patch}
\end{align}
for the correlation function, where we exploited in the last step that $G_0$ 
and $\sigma_z$ commute. Rotation of the Fermi momentum, with respect to which 
the patch theory is defined, by a small angle $\theta$ yields
\begin{align}
\langle T_{xy} T_{xy}\rangle_\text{1Loop}(q) = -N \intm{k} \sum_s 
\Bigl(k_y + s \frac{\theta}{2}\Bigr)^2 
G_{0,s}(k_0 + \omega, \bs k) G_{0,s}(k_0, \bs k),
\end{align}
where
\begin{align}
	k_x &\rightarrow k_x - \theta k_y - s \frac{\theta^2}{4},		&		k_y 
&\rightarrow k_y + s \frac{\theta}{2}.
\end{align}
The Green's functions are independent of $\theta$ due to the emergent rotation 
symmetry. We can therefore eliminate $k_y$ from the Green's functions by 
shifting $k_x \rightarrow k_x - s k_y^2$. Then $\theta$ only appears in the 
integrand of the $k_y$ integral, which is just a multiplicative prefactor, and 
can be eliminated by shifting $k_y \rightarrow k_y - s \frac{\theta}{2}$. 
Nearby patches thus contribute equally to the $T_{xy}$ correlator.

This result can be complemented by an analysis of the stress tensor 
correlation function for a continuum model. For an isotropic system we can 
start 
from the Lagrangian
\begin{equation}
	\mathcal L(x) = \psi^\dagger(x) \partial_\tau \psi(x) + 
\bs\nabla\psi^\dagger(x) \cdot \bs \nabla \psi(x) - \mu \psi^\dagger(x) \psi(x),
\end{equation}
where we omitted the interaction and the bosonic contribution. The 
$xy$-component of the stress tensor reads
\begin{equation}
	T_{xy}(q) = \intm{k} \bigl((k_x + q_x) k_y + (k_y + q_y) k_x\bigr) 
\psi^\dagger(k+q) \psi(k).
\end{equation}
At one-loop level, the $T_{xy}$ autocorrelation function for $\bs q = 0$ is 
then given by
\begin{equation}
	\langle T_{xy} T_{xy}\rangle_\text{1Loop}(i\omega) = -4 \intm{k} k_x^2 
k_y^2 G_0(k+q) G_0(k)
\label{eq:TxyTxy_1loop_FS_Continuum}
\end{equation}
where $q = \omega \bs e_0$.

We can subdivide the vicinity of the Fermi surface into (finite) patches, which 
are labeled by $\phi$, and obtain
\begin{align}
\langle T_{xy} T_{xy} \rangle_\text{1Loop}(i\omega) = -4 \sum_\text{Patches} 
\intm{k'} (k_x' \cos \phi - k_y' \sin \phi)^2  (k_x' \sin \phi + k_y' \cos 
\phi)^2 \operatorname{tr}\bigl(G_0(k_0' + \omega, \bs k') G_0(k_0', \bs 
k')\bigr),
\end{align}
where the integral over $\bs k'$ is over a specific patch. The sum over patches 
(or $\phi$-integration) sums up the contributions from individual patches. The 
Green's functions do not depend on $\phi$ because they are the same in each 
local patch coordinate system (Fig.~\ref{fig:Patchrotate}) and are just given by
the patch theory action in the supplement. Shifting $k_x' \rightarrow k_F + 
k_x'$ in order to 
make the Fermi momentum explicit, we obtain
\begin{align}
(k_x' \cos \phi - k_y' \sin \phi)^2 &(k_x' \sin \phi + k_y' \cos 
\phi)^2 \rightarrow  \frac{1}{4} \left(2 k_F k_y' \cos (2 \phi )+k_F^2 \sin (2 
\phi) - {k_y'}^2 \sin (2 \phi )\right)^2 \nonumber \\
&+\frac{1}{4} {k_x'}^2 \left(6 k_F k_y' \sin (4 \phi )+6 k_F^2 \sin^2(2 \phi) + 
{k_y'}^2 (3 \cos (4 \phi )+1)\right) \nonumber \\
&+\frac{1}{2} k_x' \Big(3 k_F^2 k_y' \sin (4 \phi ) + k_F {k_y'}^2 (3 \cos(4 
\phi) + 1)  + 2 k_F^3 \sin^2(2\phi) - {k_y'}^3 \sin(4\phi)\Big) \nonumber \\
& + {k_x'}^3 \sin(2\phi) \left(k_F \sin(2\phi) + k_y' \cos(2\phi)\right) + 
{k_x'}^4 \sin ^2\phi \cos ^2\phi.
\label{eq:TxyTxy_MomentumFactor_Expanded}
\end{align}
The terms in the first line of the right hand side do not depend on 
$k_x^\prime$ 
and yield the
scaling that we determined from the patch theory as $k_y^\prime$ and $k_F$ do 
not scale.
The terms on the other lines contain additional powers of 
$k_x^\prime\sim\omega^{1/z}$ and
are hence subleading.
\begin{figure}[h]
	\centering
	\includegraphics[height=2.0in]{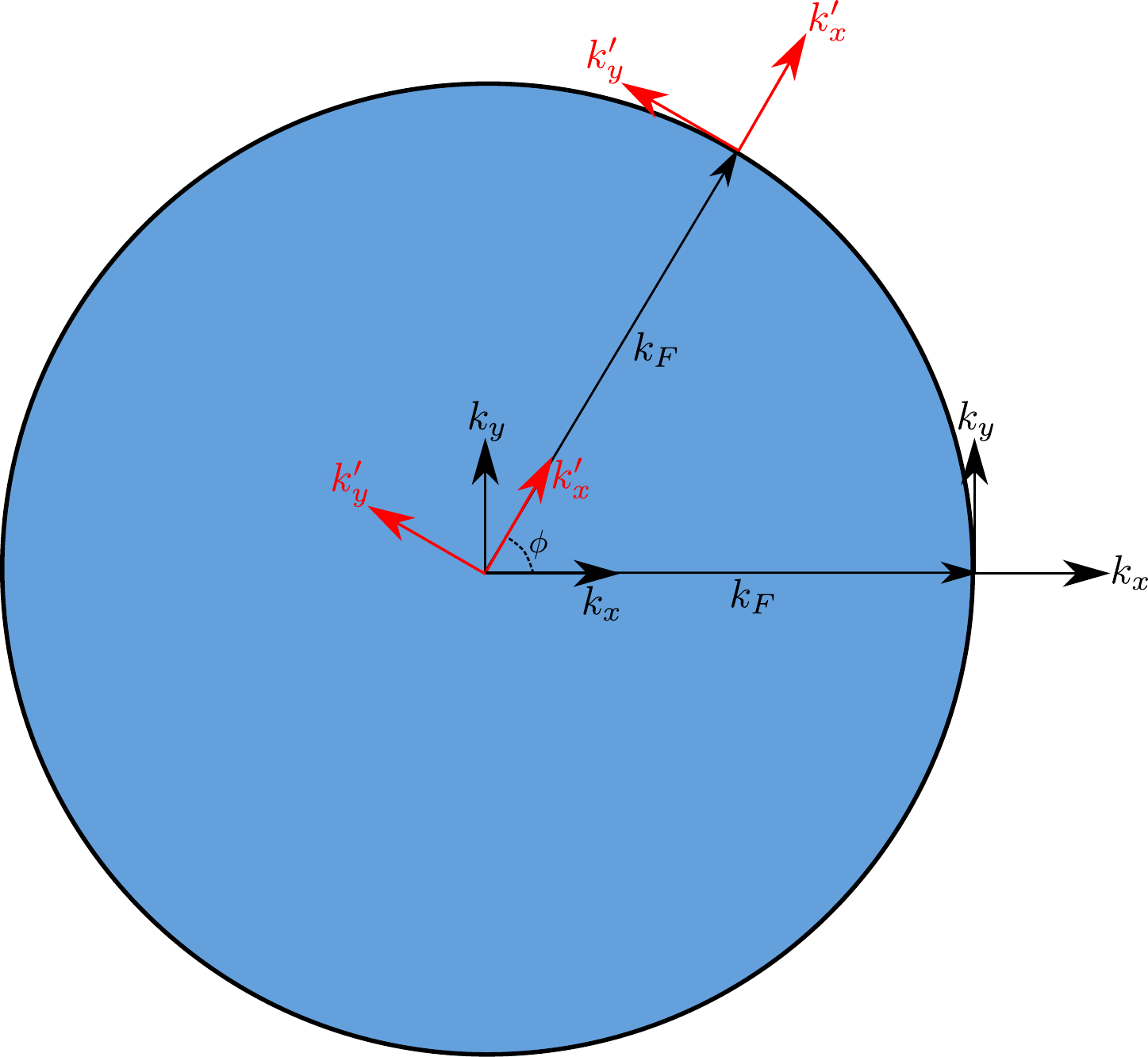}
	\caption{Transformation of coordinates used to determine the contribution of 
different patches to the $T_{xy}$~-~$T_{xy}$ correlator.}
	\label{fig:Patchrotate}
\end{figure}
The above argument takes care of the 1-loop and self-energy corrections. In the 
vertex corrections, the 
additional powers of $k_x^\prime$ and $k_y^\prime$ do not influence the absence 
of poles in $\epsilon^{-1}$. Hence all 
patches contribute the same scaling at leading order in $\omega$. 

The expressions in Eqs.~\eqref{eq:TxyTxy_1loop_Patch} 
and~\eqref{eq:TxyTxy_1loop_FS_Continuum} are directly related only for the 
patches in the $k_x$ or the $k_y$ direction. In the former case, evaluating the 
factor $k_x^2 k_y^2$ in the integrand close to the Fermi surface yields $(k_F + 
k_x')^2 {k_y'}^2 \approx k_F^2 {k_y'}^2$. After rescaling of momentum 
variables, this yields the factor of $k_y^2$ that appears in the stress tensor 
correlation function of the patch theory in Eq.~\eqref{eq:TxyTxy_1loop_Patch}. 
For other directions additional terms appear, which are not present in the 
patch 
theory, for example terms in Eq.~\eqref{eq:TxyTxy_MomentumFactor_Expanded} 
which 
are proportional to $k_F^4$. As $k_F$ and $k_y$ do not scale, such terms are 
equally relevant to the terms that appear in the patch theory and thus do not 
change the scaling behavior. The argument employing the emergent 
rotational symmetry does not generate such terms, but nevertheless leads to the 
correct scaling behavior.

\end{widetext}

\bibliography{IsingViscosity}

\begin{thebibliography}{60}%
\makeatletter
\providecommand \@ifxundefined [1]{%
 \@ifx{#1\undefined}
}%
\providecommand \@ifnum [1]{%
 \ifnum #1\expandafter \@firstoftwo
 \else \expandafter \@secondoftwo
 \fi
}%
\providecommand \@ifx [1]{%
 \ifx #1\expandafter \@firstoftwo
 \else \expandafter \@secondoftwo
 \fi
}%
\providecommand \natexlab [1]{#1}%
\providecommand \enquote  [1]{``#1''}%
\providecommand \bibnamefont  [1]{#1}%
\providecommand \bibfnamefont [1]{#1}%
\providecommand \citenamefont [1]{#1}%
\providecommand \href@noop [0]{\@secondoftwo}%
\providecommand \href [0]{\begingroup \@sanitize@url \@href}%
\providecommand \@href[1]{\@@startlink{#1}\@@href}%
\providecommand \@@href[1]{\endgroup#1\@@endlink}%
\providecommand \@sanitize@url [0]{\catcode `\\12\catcode `\$12\catcode
  `\&12\catcode `\#12\catcode `\^12\catcode `\_12\catcode `\%12\relax}%
\providecommand \@@startlink[1]{}%
\providecommand \@@endlink[0]{}%
\providecommand \url  [0]{\begingroup\@sanitize@url \@url }%
\providecommand \@url [1]{\endgroup\@href {#1}{\urlprefix }}%
\providecommand \urlprefix  [0]{URL }%
\providecommand \Eprint [0]{\href }%
\providecommand \doibase [0]{http://dx.doi.org/}%
\providecommand \selectlanguage [0]{\@gobble}%
\providecommand \bibinfo  [0]{\@secondoftwo}%
\providecommand \bibfield  [0]{\@secondoftwo}%
\providecommand \translation [1]{[#1]}%
\providecommand \BibitemOpen [0]{}%
\providecommand \bibitemStop [0]{}%
\providecommand \bibitemNoStop [0]{.\EOS\space}%
\providecommand \EOS [0]{\spacefactor3000\relax}%
\providecommand \BibitemShut  [1]{\csname bibitem#1\endcsname}%
\let\auto@bib@innerbib\@empty
\bibitem [{\citenamefont {{Bandurin}}\ \emph {et~al.}(2016)\citenamefont
  {{Bandurin}}, \citenamefont {{Torre}}, \citenamefont {{Kumar}}, \citenamefont
  {{Ben Shalom}}, \citenamefont {{Tomadin}}, \citenamefont {{Principi}},
  \citenamefont {{Auton}}, \citenamefont {{Khestanova}}, \citenamefont
  {{Novoselov}}, \citenamefont {{Grigorieva}}, \citenamefont {{Ponomarenko}},
  \citenamefont {{Geim}},\ and\ \citenamefont {{Polini}}}]{Geim16}%
  \BibitemOpen
  \bibfield  {author} {\bibinfo {author} {\bibfnamefont {D.~A.}\ \bibnamefont
  {{Bandurin}}}, \bibinfo {author} {\bibfnamefont {I.}~\bibnamefont {{Torre}}},
  \bibinfo {author} {\bibfnamefont {R.~K.}\ \bibnamefont {{Kumar}}}, \bibinfo
  {author} {\bibfnamefont {M.}~\bibnamefont {{Ben Shalom}}}, \bibinfo {author}
  {\bibfnamefont {A.}~\bibnamefont {{Tomadin}}}, \bibinfo {author}
  {\bibfnamefont {A.}~\bibnamefont {{Principi}}}, \bibinfo {author}
  {\bibfnamefont {G.~H.}\ \bibnamefont {{Auton}}}, \bibinfo {author}
  {\bibfnamefont {E.}~\bibnamefont {{Khestanova}}}, \bibinfo {author}
  {\bibfnamefont {K.~S.}\ \bibnamefont {{Novoselov}}}, \bibinfo {author}
  {\bibfnamefont {I.~V.}\ \bibnamefont {{Grigorieva}}}, \bibinfo {author}
  {\bibfnamefont {L.~A.}\ \bibnamefont {{Ponomarenko}}}, \bibinfo {author}
  {\bibfnamefont {A.~K.}\ \bibnamefont {{Geim}}}, \ and\ \bibinfo {author}
  {\bibfnamefont {M.}~\bibnamefont {{Polini}}},\ }\bibfield  {title} {\enquote
  {\bibinfo {title} {{Negative local resistance caused by viscous electron
  backflow in graphene}},}\ }\href {\doibase 10.1126/science.aad0201}
  {\bibfield  {journal} {\bibinfo  {journal} {Science}\ }\textbf {\bibinfo
  {volume} {351}},\ \bibinfo {pages} {1055} (\bibinfo {year} {2016})},\ \Eprint
  {http://arxiv.org/abs/1509.04165} {arXiv:1509.04165 [cond-mat.str-el]}
  \BibitemShut {NoStop}%
\bibitem [{\citenamefont {{Crossno}}\ \emph {et~al.}(2016)\citenamefont
  {{Crossno}}, \citenamefont {{Shi}}, \citenamefont {{Wang}}, \citenamefont
  {{Liu}}, \citenamefont {{Harzheim}}, \citenamefont {{Lucas}}, \citenamefont
  {{Sachdev}}, \citenamefont {{Kim}}, \citenamefont {{Taniguchi}},
  \citenamefont {{Watanabe}}, \citenamefont {{Ohki}},\ and\ \citenamefont
  {{Fong}}}]{Kim16}%
  \BibitemOpen
  \bibfield  {author} {\bibinfo {author} {\bibfnamefont {J.}~\bibnamefont
  {{Crossno}}}, \bibinfo {author} {\bibfnamefont {J.~K.}\ \bibnamefont
  {{Shi}}}, \bibinfo {author} {\bibfnamefont {K.}~\bibnamefont {{Wang}}},
  \bibinfo {author} {\bibfnamefont {X.}~\bibnamefont {{Liu}}}, \bibinfo
  {author} {\bibfnamefont {A.}~\bibnamefont {{Harzheim}}}, \bibinfo {author}
  {\bibfnamefont {A.}~\bibnamefont {{Lucas}}}, \bibinfo {author} {\bibfnamefont
  {S.}~\bibnamefont {{Sachdev}}}, \bibinfo {author} {\bibfnamefont
  {P.}~\bibnamefont {{Kim}}}, \bibinfo {author} {\bibfnamefont
  {T.}~\bibnamefont {{Taniguchi}}}, \bibinfo {author} {\bibfnamefont
  {K.}~\bibnamefont {{Watanabe}}}, \bibinfo {author} {\bibfnamefont {T.~A.}\
  \bibnamefont {{Ohki}}}, \ and\ \bibinfo {author} {\bibfnamefont {K.~C.}\
  \bibnamefont {{Fong}}},\ }\bibfield  {title} {\enquote {\bibinfo {title}
  {{Observation of the Dirac fluid and the breakdown of the Wiedemann-Franz law
  in graphene}},}\ }\href {\doibase 10.1126/science.aad0343} {\bibfield
  {journal} {\bibinfo  {journal} {Science}\ }\textbf {\bibinfo {volume}
  {351}},\ \bibinfo {pages} {1058} (\bibinfo {year} {2016})},\ \Eprint
  {http://arxiv.org/abs/1509.04713} {arXiv:1509.04713 [cond-mat.mes-hall]}
  \BibitemShut {NoStop}%
\bibitem [{\citenamefont {{Moll}}\ \emph {et~al.}(2016)\citenamefont {{Moll}},
  \citenamefont {{Kushwaha}}, \citenamefont {{Nandi}}, \citenamefont
  {{Schmidt}},\ and\ \citenamefont {{Mackenzie}}}]{APM16}%
  \BibitemOpen
  \bibfield  {author} {\bibinfo {author} {\bibfnamefont {P.~J.~W.}\
  \bibnamefont {{Moll}}}, \bibinfo {author} {\bibfnamefont {P.}~\bibnamefont
  {{Kushwaha}}}, \bibinfo {author} {\bibfnamefont {N.}~\bibnamefont {{Nandi}}},
  \bibinfo {author} {\bibfnamefont {B.}~\bibnamefont {{Schmidt}}}, \ and\
  \bibinfo {author} {\bibfnamefont {A.~P.}\ \bibnamefont {{Mackenzie}}},\
  }\bibfield  {title} {\enquote {\bibinfo {title} {{Evidence for hydrodynamic
  electron flow in PdCoO$_{2}$}},}\ }\href {\doibase 10.1126/science.aac8385}
  {\bibfield  {journal} {\bibinfo  {journal} {Science}\ }\textbf {\bibinfo
  {volume} {351}},\ \bibinfo {pages} {1061} (\bibinfo {year} {2016})},\ \Eprint
  {http://arxiv.org/abs/1509.05691} {arXiv:1509.05691 [cond-mat.str-el]}
  \BibitemShut {NoStop}%
\bibitem [{\citenamefont {Gurzhi}(1968)}]{Gurzhi}%
  \BibitemOpen
  \bibfield  {author} {\bibinfo {author} {\bibfnamefont {R.~N.}\ \bibnamefont
  {Gurzhi}},\ }\bibfield  {title} {\enquote {\bibinfo {title} {{Hydrodynamic
  effects in solids at low temperatures}},}\ }\href
  {http://stacks.iop.org/0038-5670/11/i=2/a=R07} {\bibfield  {journal}
  {\bibinfo  {journal} {Sov. Phys. Usp.}\ }\textbf {\bibinfo {volume} {11}},\
  \bibinfo {pages} {255} (\bibinfo {year} {1968})}\BibitemShut {NoStop}%
\bibitem [{\citenamefont {{de Jong}}\ and\ \citenamefont
  {{Molenkamp}}(1995)}]{Molen95}%
  \BibitemOpen
  \bibfield  {author} {\bibinfo {author} {\bibfnamefont {M.~J.~M.}\
  \bibnamefont {{de Jong}}}\ and\ \bibinfo {author} {\bibfnamefont {L.~W.}\
  \bibnamefont {{Molenkamp}}},\ }\bibfield  {title} {\enquote {\bibinfo {title}
  {{Hydrodynamic electron flow in high-mobility wires}},}\ }\href {\doibase
  10.1103/PhysRevB.51.13389} {\bibfield  {journal} {\bibinfo  {journal} {\prb}\
  }\textbf {\bibinfo {volume} {51}},\ \bibinfo {pages} {13389} (\bibinfo {year}
  {1995})},\ \Eprint {http://arxiv.org/abs/cond-mat/9411067} {cond-mat/9411067}
  \BibitemShut {NoStop}%
\bibitem [{\citenamefont {{Andreev}}\ \emph {et~al.}(2011)\citenamefont
  {{Andreev}}, \citenamefont {{Kivelson}},\ and\ \citenamefont
  {{Spivak}}}]{AKS10}%
  \BibitemOpen
  \bibfield  {author} {\bibinfo {author} {\bibfnamefont {A.~V.}\ \bibnamefont
  {{Andreev}}}, \bibinfo {author} {\bibfnamefont {S.~A.}\ \bibnamefont
  {{Kivelson}}}, \ and\ \bibinfo {author} {\bibfnamefont {B.}~\bibnamefont
  {{Spivak}}},\ }\bibfield  {title} {\enquote {\bibinfo {title} {{Hydrodynamic
  Description of Transport in Strongly Correlated Electron Systems}},}\ }\href
  {\doibase 10.1103/PhysRevLett.106.256804} {\bibfield  {journal} {\bibinfo
  {journal} {Phys. Rev. Lett.}\ }\textbf {\bibinfo {volume} {106}},\ \bibinfo
  {eid} {256804} (\bibinfo {year} {2011})},\ \Eprint
  {http://arxiv.org/abs/1011.3068} {arXiv:1011.3068 [cond-mat.mes-hall]}
  \BibitemShut {NoStop}%
\bibitem [{\citenamefont {{Damle}}\ and\ \citenamefont
  {{Sachdev}}(1997)}]{Damle97}%
  \BibitemOpen
  \bibfield  {author} {\bibinfo {author} {\bibfnamefont {K.}~\bibnamefont
  {{Damle}}}\ and\ \bibinfo {author} {\bibfnamefont {S.}~\bibnamefont
  {{Sachdev}}},\ }\bibfield  {title} {\enquote {\bibinfo {title}
  {{Nonzero-temperature transport near quantum critical points}},}\ }\href
  {\doibase 10.1103/PhysRevB.56.8714} {\bibfield  {journal} {\bibinfo
  {journal} {Phys. Rev. B}\ }\textbf {\bibinfo {volume} {56}},\ \bibinfo
  {pages} {8714} (\bibinfo {year} {1997})},\ \Eprint
  {http://arxiv.org/abs/cond-mat/9705206} {cond-mat/9705206} \BibitemShut
  {NoStop}%
\bibitem [{\citenamefont {{Hartnoll}}\ \emph {et~al.}(2007)\citenamefont
  {{Hartnoll}}, \citenamefont {{Kovtun}}, \citenamefont {{M{\"u}ller}},\ and\
  \citenamefont {{Sachdev}}}]{HKMS}%
  \BibitemOpen
  \bibfield  {author} {\bibinfo {author} {\bibfnamefont {S.~A.}\ \bibnamefont
  {{Hartnoll}}}, \bibinfo {author} {\bibfnamefont {P.~K.}\ \bibnamefont
  {{Kovtun}}}, \bibinfo {author} {\bibfnamefont {M.}~\bibnamefont
  {{M{\"u}ller}}}, \ and\ \bibinfo {author} {\bibfnamefont {S.}~\bibnamefont
  {{Sachdev}}},\ }\bibfield  {title} {\enquote {\bibinfo {title} {{Theory of
  the Nernst effect near quantum phase transitions in condensed matter and in
  dyonic black holes}},}\ }\href {\doibase 10.1103/PhysRevB.76.144502}
  {\bibfield  {journal} {\bibinfo  {journal} {Phys. Rev. B}\ }\textbf {\bibinfo
  {volume} {76}},\ \bibinfo {eid} {144502} (\bibinfo {year} {2007})},\ \Eprint
  {http://arxiv.org/abs/0706.3215} {arXiv:0706.3215 [cond-mat.str-el]}
  \BibitemShut {NoStop}%
\bibitem [{\citenamefont {{Hartnoll}}\ \emph {et~al.}(2014)\citenamefont
  {{Hartnoll}}, \citenamefont {{Mahajan}}, \citenamefont {{Punk}},\ and\
  \citenamefont {{Sachdev}}}]{HMPS14}%
  \BibitemOpen
  \bibfield  {author} {\bibinfo {author} {\bibfnamefont {S.~A.}\ \bibnamefont
  {{Hartnoll}}}, \bibinfo {author} {\bibfnamefont {R.}~\bibnamefont
  {{Mahajan}}}, \bibinfo {author} {\bibfnamefont {M.}~\bibnamefont {{Punk}}}, \
  and\ \bibinfo {author} {\bibfnamefont {S.}~\bibnamefont {{Sachdev}}},\
  }\bibfield  {title} {\enquote {\bibinfo {title} {{Transport near the
  Ising-nematic quantum critical point of metals in two dimensions}},}\ }\href
  {\doibase 10.1103/PhysRevB.89.155130} {\bibfield  {journal} {\bibinfo
  {journal} {Phys. Rev. B}\ }\textbf {\bibinfo {volume} {89}},\ \bibinfo {eid}
  {155130} (\bibinfo {year} {2014})},\ \Eprint {http://arxiv.org/abs/1401.7012}
  {arXiv:1401.7012 [cond-mat.str-el]} \BibitemShut {NoStop}%
\bibitem [{\citenamefont {{Patel}}\ and\ \citenamefont
  {{Sachdev}}(2014)}]{PS14}%
  \BibitemOpen
  \bibfield  {author} {\bibinfo {author} {\bibfnamefont {A.~A.}\ \bibnamefont
  {{Patel}}}\ and\ \bibinfo {author} {\bibfnamefont {S.}~\bibnamefont
  {{Sachdev}}},\ }\bibfield  {title} {\enquote {\bibinfo {title} {{DC}
  resistivity at the onset of spin density wave order in two-dimensional
  metals},}\ }\href {\doibase 10.1103/PhysRevB.90.165146} {\bibfield  {journal}
  {\bibinfo  {journal} {Phys. Rev. B}\ }\textbf {\bibinfo {volume} {90}},\
  \bibinfo {eid} {165146} (\bibinfo {year} {2014})},\ \Eprint
  {http://arxiv.org/abs/1408.6549} {arXiv:1408.6549 [cond-mat.str-el]}
  \BibitemShut {NoStop}%
\bibitem [{\citenamefont {{Lucas}}\ and\ \citenamefont
  {{Sachdev}}(2015)}]{LS15}%
  \BibitemOpen
  \bibfield  {author} {\bibinfo {author} {\bibfnamefont {A.}~\bibnamefont
  {{Lucas}}}\ and\ \bibinfo {author} {\bibfnamefont {S.}~\bibnamefont
  {{Sachdev}}},\ }\bibfield  {title} {\enquote {\bibinfo {title} {{Memory
  matrix theory of magnetotransport in strange metals}},}\ }\href {\doibase
  10.1103/PhysRevB.91.195122} {\bibfield  {journal} {\bibinfo  {journal} {Phys.
  Rev. B}\ }\textbf {\bibinfo {volume} {91}},\ \bibinfo {eid} {195122}
  (\bibinfo {year} {2015})},\ \Eprint {http://arxiv.org/abs/1502.04704}
  {arXiv:1502.04704 [cond-mat.str-el]} \BibitemShut {NoStop}%
\bibitem [{\citenamefont {{M{\"u}ller}}\ and\ \citenamefont
  {{Sachdev}}(2008)}]{MMSS08}%
  \BibitemOpen
  \bibfield  {author} {\bibinfo {author} {\bibfnamefont {M.}~\bibnamefont
  {{M{\"u}ller}}}\ and\ \bibinfo {author} {\bibfnamefont {S.}~\bibnamefont
  {{Sachdev}}},\ }\bibfield  {title} {\enquote {\bibinfo {title} {{Collective
  cyclotron motion of the relativistic plasma in graphene}},}\ }\href {\doibase
  10.1103/PhysRevB.78.115419} {\bibfield  {journal} {\bibinfo  {journal} {Phys.
  Rev. B}\ }\textbf {\bibinfo {volume} {78}},\ \bibinfo {eid} {115419}
  (\bibinfo {year} {2008})},\ \Eprint {http://arxiv.org/abs/0801.2970}
  {arXiv:0801.2970 [cond-mat.str-el]} \BibitemShut {NoStop}%
\bibitem [{\citenamefont {{Fritz}}\ \emph {et~al.}(2008)\citenamefont
  {{Fritz}}, \citenamefont {{Schmalian}}, \citenamefont {{M{\"u}ller}},\ and\
  \citenamefont {{Sachdev}}}]{FSMS08}%
  \BibitemOpen
  \bibfield  {author} {\bibinfo {author} {\bibfnamefont {L.}~\bibnamefont
  {{Fritz}}}, \bibinfo {author} {\bibfnamefont {J.}~\bibnamefont
  {{Schmalian}}}, \bibinfo {author} {\bibfnamefont {M.}~\bibnamefont
  {{M{\"u}ller}}}, \ and\ \bibinfo {author} {\bibfnamefont {S.}~\bibnamefont
  {{Sachdev}}},\ }\bibfield  {title} {\enquote {\bibinfo {title} {{Quantum
  critical transport in clean graphene}},}\ }\href {\doibase
  10.1103/PhysRevB.78.085416} {\bibfield  {journal} {\bibinfo  {journal} {Phys.
  Rev. B}\ }\textbf {\bibinfo {volume} {78}},\ \bibinfo {eid} {085416}
  (\bibinfo {year} {2008})},\ \Eprint {http://arxiv.org/abs/0802.4289}
  {arXiv:0802.4289} \BibitemShut {NoStop}%
\bibitem [{\citenamefont {{M{\"u}ller}}\ \emph {et~al.}(2008)\citenamefont
  {{M{\"u}ller}}, \citenamefont {{Fritz}},\ and\ \citenamefont
  {{Sachdev}}}]{MFS08}%
  \BibitemOpen
  \bibfield  {author} {\bibinfo {author} {\bibfnamefont {M.}~\bibnamefont
  {{M{\"u}ller}}}, \bibinfo {author} {\bibfnamefont {L.}~\bibnamefont
  {{Fritz}}}, \ and\ \bibinfo {author} {\bibfnamefont {S.}~\bibnamefont
  {{Sachdev}}},\ }\bibfield  {title} {\enquote {\bibinfo {title}
  {{Quantum-critical relativistic magnetotransport in graphene}},}\ }\href
  {\doibase 10.1103/PhysRevB.78.115406} {\bibfield  {journal} {\bibinfo
  {journal} {Phys. Rev. B}\ }\textbf {\bibinfo {volume} {78}},\ \bibinfo {eid}
  {115406} (\bibinfo {year} {2008})},\ \Eprint {http://arxiv.org/abs/0805.1413}
  {arXiv:0805.1413 [cond-mat.str-el]} \BibitemShut {NoStop}%
\bibitem [{\citenamefont {{Foster}}\ and\ \citenamefont
  {{Aleiner}}(2009)}]{FA09}%
  \BibitemOpen
  \bibfield  {author} {\bibinfo {author} {\bibfnamefont {M.~S.}\ \bibnamefont
  {{Foster}}}\ and\ \bibinfo {author} {\bibfnamefont {I.~L.}\ \bibnamefont
  {{Aleiner}}},\ }\bibfield  {title} {\enquote {\bibinfo {title} {{Slow
  imbalance relaxation and thermoelectric transport in graphene}},}\ }\href
  {\doibase 10.1103/PhysRevB.79.085415} {\bibfield  {journal} {\bibinfo
  {journal} {Phys. Rev. B}\ }\textbf {\bibinfo {volume} {79}},\ \bibinfo {eid}
  {085415} (\bibinfo {year} {2009})},\ \Eprint {http://arxiv.org/abs/0810.4342}
  {arXiv:0810.4342 [cond-mat.mes-hall]} \BibitemShut {NoStop}%
\bibitem [{\citenamefont {M\"uller}\ \emph {et~al.}(2009)\citenamefont
  {M\"uller}, \citenamefont {Schmalian},\ and\ \citenamefont {Fritz}}]{MSF09}%
  \BibitemOpen
  \bibfield  {author} {\bibinfo {author} {\bibfnamefont {M.}~\bibnamefont
  {M\"uller}}, \bibinfo {author} {\bibfnamefont {J.}~\bibnamefont {Schmalian}},
  \ and\ \bibinfo {author} {\bibfnamefont {L.}~\bibnamefont {Fritz}},\
  }\bibfield  {title} {\enquote {\bibinfo {title} {{Graphene: A Nearly Perfect
  Fluid}},}\ }\href {\doibase 10.1103/PhysRevLett.103.025301} {\bibfield
  {journal} {\bibinfo  {journal} {Phys. Rev. Lett.}\ }\textbf {\bibinfo
  {volume} {103}},\ \bibinfo {pages} {025301} (\bibinfo {year} {2009})},\
  \Eprint {http://arxiv.org/abs/0903.4178} {arXiv:0903.4178
  [cond-mat.mes-hall]} \BibitemShut {NoStop}%
\bibitem [{\citenamefont {Mendoza}\ \emph {et~al.}(2011)\citenamefont
  {Mendoza}, \citenamefont {Herrmann},\ and\ \citenamefont
  {Succi}}]{Mendoza11}%
  \BibitemOpen
  \bibfield  {author} {\bibinfo {author} {\bibfnamefont {M.}~\bibnamefont
  {Mendoza}}, \bibinfo {author} {\bibfnamefont {H.~J.}\ \bibnamefont
  {Herrmann}}, \ and\ \bibinfo {author} {\bibfnamefont {S.}~\bibnamefont
  {Succi}},\ }\bibfield  {title} {\enquote {\bibinfo {title} {{Preturbulent
  Regimes in Graphene Flow}},}\ }\href {\doibase
  10.1103/PhysRevLett.106.156601} {\bibfield  {journal} {\bibinfo  {journal}
  {Phys. Rev. Lett.}\ }\textbf {\bibinfo {volume} {106}},\ \bibinfo {pages}
  {156601} (\bibinfo {year} {2011})},\ \Eprint {http://arxiv.org/abs/1201.6590}
  {arXiv:1201.6590 [cond-mat.mes-hall]} \BibitemShut {NoStop}%
\bibitem [{\citenamefont {{Tomadin}}\ \emph {et~al.}(2014)\citenamefont
  {{Tomadin}}, \citenamefont {{Vignale}},\ and\ \citenamefont
  {{Polini}}}]{Polini14}%
  \BibitemOpen
  \bibfield  {author} {\bibinfo {author} {\bibfnamefont {A.}~\bibnamefont
  {{Tomadin}}}, \bibinfo {author} {\bibfnamefont {G.}~\bibnamefont
  {{Vignale}}}, \ and\ \bibinfo {author} {\bibfnamefont {M.}~\bibnamefont
  {{Polini}}},\ }\bibfield  {title} {\enquote {\bibinfo {title} {{Corbino Disk
  Viscometer for 2D Quantum Electron Liquids}},}\ }\href {\doibase
  10.1103/PhysRevLett.113.235901} {\bibfield  {journal} {\bibinfo  {journal}
  {Phys. Rev. Lett.}\ }\textbf {\bibinfo {volume} {113}},\ \bibinfo {eid}
  {235901} (\bibinfo {year} {2014})},\ \Eprint {http://arxiv.org/abs/1401.0938}
  {arXiv:1401.0938 [cond-mat.mes-hall]} \BibitemShut {NoStop}%
\bibitem [{\citenamefont {{Principi}}\ and\ \citenamefont
  {{Vignale}}(2015)}]{Vignale15}%
  \BibitemOpen
  \bibfield  {author} {\bibinfo {author} {\bibfnamefont {A.}~\bibnamefont
  {{Principi}}}\ and\ \bibinfo {author} {\bibfnamefont {G.}~\bibnamefont
  {{Vignale}}},\ }\bibfield  {title} {\enquote {\bibinfo {title} {{Violation of
  the Wiedemann-Franz Law in Hydrodynamic Electron Liquids}},}\ }\href
  {\doibase 10.1103/PhysRevLett.115.056603} {\bibfield  {journal} {\bibinfo
  {journal} {Phys. Rev. Lett.}\ }\textbf {\bibinfo {volume} {115}},\ \bibinfo
  {eid} {056603} (\bibinfo {year} {2015})},\ \Eprint
  {http://arxiv.org/abs/1406.2940} {arXiv:1406.2940 [cond-mat.mes-hall]}
  \BibitemShut {NoStop}%
\bibitem [{\citenamefont {{Torre}}\ \emph {et~al.}(2015)\citenamefont
  {{Torre}}, \citenamefont {{Tomadin}}, \citenamefont {{Geim}},\ and\
  \citenamefont {{Polini}}}]{Polini15}%
  \BibitemOpen
  \bibfield  {author} {\bibinfo {author} {\bibfnamefont {I.}~\bibnamefont
  {{Torre}}}, \bibinfo {author} {\bibfnamefont {A.}~\bibnamefont {{Tomadin}}},
  \bibinfo {author} {\bibfnamefont {A.~K.}\ \bibnamefont {{Geim}}}, \ and\
  \bibinfo {author} {\bibfnamefont {M.}~\bibnamefont {{Polini}}},\ }\bibfield
  {title} {\enquote {\bibinfo {title} {{Nonlocal transport and the hydrodynamic
  shear viscosity in graphene}},}\ }\href {\doibase 10.1103/PhysRevB.92.165433}
  {\bibfield  {journal} {\bibinfo  {journal} {Phys. Rev. B}\ }\textbf {\bibinfo
  {volume} {92}},\ \bibinfo {eid} {165433} (\bibinfo {year} {2015})},\ \Eprint
  {http://arxiv.org/abs/1508.00363} {arXiv:1508.00363 [cond-mat.mes-hall]}
  \BibitemShut {NoStop}%
\bibitem [{\citenamefont {Levitov}\ and\ \citenamefont
  {Falkovich}(2016)}]{Levitov16}%
  \BibitemOpen
  \bibfield  {author} {\bibinfo {author} {\bibfnamefont {L.}~\bibnamefont
  {Levitov}}\ and\ \bibinfo {author} {\bibfnamefont {G.}~\bibnamefont
  {Falkovich}},\ }\bibfield  {title} {\enquote {\bibinfo {title} {{Electron
  viscosity, current vortices and negative nonlocal resistance in graphene}},}\
  }\href {http://dx.doi.org/10.1038/nphys3667} {\bibfield  {journal} {\bibinfo
  {journal} {Nat Phys}\ }\textbf {\bibinfo {volume} {12}},\ \bibinfo {pages}
  {672} (\bibinfo {year} {2016})},\ \Eprint {http://arxiv.org/abs/1508.00836}
  {arXiv:1508.00836 [cond-mat.mes-hall]} \BibitemShut {NoStop}%
\bibitem [{\citenamefont {{Lucas}}\ \emph {et~al.}(2016)\citenamefont
  {{Lucas}}, \citenamefont {{Crossno}}, \citenamefont {{Fong}}, \citenamefont
  {{Kim}},\ and\ \citenamefont {{Sachdev}}}]{ALJC15}%
  \BibitemOpen
  \bibfield  {author} {\bibinfo {author} {\bibfnamefont {A.}~\bibnamefont
  {{Lucas}}}, \bibinfo {author} {\bibfnamefont {J.}~\bibnamefont {{Crossno}}},
  \bibinfo {author} {\bibfnamefont {K.~C.}\ \bibnamefont {{Fong}}}, \bibinfo
  {author} {\bibfnamefont {P.}~\bibnamefont {{Kim}}}, \ and\ \bibinfo {author}
  {\bibfnamefont {S.}~\bibnamefont {{Sachdev}}},\ }\bibfield  {title} {\enquote
  {\bibinfo {title} {{Transport in inhomogeneous quantum critical fluids and in
  the Dirac fluid in graphene}},}\ }\href {\doibase 10.1103/PhysRevB.93.075426}
  {\bibfield  {journal} {\bibinfo  {journal} {Phys. Rev. B}\ }\textbf {\bibinfo
  {volume} {93}},\ \bibinfo {eid} {075426} (\bibinfo {year} {2016})},\ \Eprint
  {http://arxiv.org/abs/1510.01738} {arXiv:1510.01738 [cond-mat.str-el]}
  \BibitemShut {NoStop}%
\bibitem [{\citenamefont {{Falkovich}}\ and\ \citenamefont
  {{Levitov}}(2016)}]{Levitov16b}%
  \BibitemOpen
  \bibfield  {author} {\bibinfo {author} {\bibfnamefont {G.}~\bibnamefont
  {{Falkovich}}}\ and\ \bibinfo {author} {\bibfnamefont {L.}~\bibnamefont
  {{Levitov}}},\ }\bibfield  {title} {\enquote {\bibinfo {title} {{Linking
  Spatial Distributions of Potential and Current in Viscous Electronics}},}\
  }\href@noop {} {\bibfield  {journal} {\bibinfo  {journal} {ArXiv e-prints}\ }
  (\bibinfo {year} {2016})},\ \Eprint {http://arxiv.org/abs/1607.00986}
  {arXiv:1607.00986 [cond-mat.mes-hall]} \BibitemShut {NoStop}%
\bibitem [{\citenamefont {{Policastro}}\ \emph {et~al.}(2001)\citenamefont
  {{Policastro}}, \citenamefont {{Son}},\ and\ \citenamefont
  {{Starinets}}}]{PSS01}%
  \BibitemOpen
  \bibfield  {author} {\bibinfo {author} {\bibfnamefont {G.}~\bibnamefont
  {{Policastro}}}, \bibinfo {author} {\bibfnamefont {D.~T.}\ \bibnamefont
  {{Son}}}, \ and\ \bibinfo {author} {\bibfnamefont {A.~O.}\ \bibnamefont
  {{Starinets}}},\ }\bibfield  {title} {\enquote {\bibinfo {title} {{Shear
  Viscosity of Strongly Coupled N = 4 Supersymmetric Yang-Mills Plasma}},}\
  }\href {\doibase 10.1103/PhysRevLett.87.081601} {\bibfield  {journal}
  {\bibinfo  {journal} {Phys. Rev. Lett.}\ }\textbf {\bibinfo {volume} {87}},\
  \bibinfo {eid} {081601} (\bibinfo {year} {2001})},\ \Eprint
  {http://arxiv.org/abs/hep-th/0104066} {hep-th/0104066} \BibitemShut {NoStop}%
\bibitem [{\citenamefont {{Kovtun}}\ \emph {et~al.}(2005)\citenamefont
  {{Kovtun}}, \citenamefont {{Son}},\ and\ \citenamefont
  {{Starinets}}}]{KSS05}%
  \BibitemOpen
  \bibfield  {author} {\bibinfo {author} {\bibfnamefont {P.~K.}\ \bibnamefont
  {{Kovtun}}}, \bibinfo {author} {\bibfnamefont {D.~T.}\ \bibnamefont {{Son}}},
  \ and\ \bibinfo {author} {\bibfnamefont {A.~O.}\ \bibnamefont
  {{Starinets}}},\ }\bibfield  {title} {\enquote {\bibinfo {title} {{Viscosity
  in Strongly Interacting Quantum Field Theories from Black Hole Physics}},}\
  }\href {\doibase 10.1103/PhysRevLett.94.111601} {\bibfield  {journal}
  {\bibinfo  {journal} {Phys. Rev. Lett.}\ }\textbf {\bibinfo {volume} {94}},\
  \bibinfo {eid} {111601} (\bibinfo {year} {2005})},\ \Eprint
  {http://arxiv.org/abs/hep-th/0405231} {hep-th/0405231} \BibitemShut {NoStop}%
\bibitem [{\citenamefont {{Karsch}}\ \emph {et~al.}(2008)\citenamefont
  {{Karsch}}, \citenamefont {{Kharzeev}},\ and\ \citenamefont
  {{Tuchin}}}]{KKT08}%
  \BibitemOpen
  \bibfield  {author} {\bibinfo {author} {\bibfnamefont {F.}~\bibnamefont
  {{Karsch}}}, \bibinfo {author} {\bibfnamefont {D.}~\bibnamefont
  {{Kharzeev}}}, \ and\ \bibinfo {author} {\bibfnamefont {K.}~\bibnamefont
  {{Tuchin}}},\ }\bibfield  {title} {\enquote {\bibinfo {title} {{Universal
  properties of bulk viscosity near the QCD phase transition}},}\ }\href
  {\doibase 10.1016/j.physletb.2008.01.080} {\bibfield  {journal} {\bibinfo
  {journal} {Phys. Lett. B}\ }\textbf {\bibinfo {volume} {663}},\ \bibinfo
  {pages} {217} (\bibinfo {year} {2008})},\ \Eprint
  {http://arxiv.org/abs/0711.0914} {arXiv:0711.0914 [hep-ph]} \BibitemShut
  {NoStop}%
\bibitem [{\citenamefont {{Heinz}}\ \emph {et~al.}(2012)\citenamefont
  {{Heinz}}, \citenamefont {{Shen}},\ and\ \citenamefont {{Song}}}]{HSS12}%
  \BibitemOpen
  \bibfield  {author} {\bibinfo {author} {\bibfnamefont {U.}~\bibnamefont
  {{Heinz}}}, \bibinfo {author} {\bibfnamefont {C.}~\bibnamefont {{Shen}}}, \
  and\ \bibinfo {author} {\bibfnamefont {H.}~\bibnamefont {{Song}}},\
  }\bibfield  {title} {\enquote {\bibinfo {title} {{The viscosity of
  quark-gluon plasma at RHIC and the LHC}},}\ }in\ \href {\doibase
  10.1063/1.3700674} {\emph {\bibinfo {booktitle} {American Institute of
  Physics Conference Series}}},\ \bibinfo {series} {American Institute of
  Physics Conference Series}, Vol.\ \bibinfo {volume} {1441},\ \bibinfo
  {editor} {edited by\ \bibinfo {editor} {\bibfnamefont {S.~G.}\ \bibnamefont
  {{Steadman}}}, \bibinfo {editor} {\bibfnamefont {G.~S.~F.}\ \bibnamefont
  {{Stephans}}}, \ and\ \bibinfo {editor} {\bibfnamefont {F.~E.}\ \bibnamefont
  {{Taylor}}}}\ (\bibinfo {year} {2012})\ pp.\ \bibinfo {pages} {766--770},\
  \Eprint {http://arxiv.org/abs/1108.5323} {arXiv:1108.5323 [nucl-th]}
  \BibitemShut {NoStop}%
\bibitem [{\citenamefont {{Cao}}\ \emph {et~al.}(2011)\citenamefont {{Cao}},
  \citenamefont {{Elliott}}, \citenamefont {{Joseph}}, \citenamefont {{Wu}},
  \citenamefont {{Petricka}}, \citenamefont {{Sch{\"a}fer}},\ and\
  \citenamefont {{Thomas}}}]{Thomas11}%
  \BibitemOpen
  \bibfield  {author} {\bibinfo {author} {\bibfnamefont {C.}~\bibnamefont
  {{Cao}}}, \bibinfo {author} {\bibfnamefont {E.}~\bibnamefont {{Elliott}}},
  \bibinfo {author} {\bibfnamefont {J.}~\bibnamefont {{Joseph}}}, \bibinfo
  {author} {\bibfnamefont {H.}~\bibnamefont {{Wu}}}, \bibinfo {author}
  {\bibfnamefont {J.}~\bibnamefont {{Petricka}}}, \bibinfo {author}
  {\bibfnamefont {T.}~\bibnamefont {{Sch{\"a}fer}}}, \ and\ \bibinfo {author}
  {\bibfnamefont {J.~E.}\ \bibnamefont {{Thomas}}},\ }\bibfield  {title}
  {\enquote {\bibinfo {title} {{Universal Quantum Viscosity in a Unitary Fermi
  Gas}},}\ }\href {\doibase 10.1126/science.1195219} {\bibfield  {journal}
  {\bibinfo  {journal} {Science}\ }\textbf {\bibinfo {volume} {331}},\ \bibinfo
  {pages} {58} (\bibinfo {year} {2011})},\ \Eprint
  {http://arxiv.org/abs/1007.2625} {arXiv:1007.2625 [cond-mat.quant-gas]}
  \BibitemShut {NoStop}%
\bibitem [{\citenamefont {{Taylor}}\ and\ \citenamefont
  {{Randeria}}(2010)}]{Taylor2010}%
  \BibitemOpen
  \bibfield  {author} {\bibinfo {author} {\bibfnamefont {E.}~\bibnamefont
  {{Taylor}}}\ and\ \bibinfo {author} {\bibfnamefont {M.}~\bibnamefont
  {{Randeria}}},\ }\bibfield  {title} {\enquote {\bibinfo {title} {{Viscosity
  of strongly interacting quantum fluids: Spectral functions and sum rules}},}\
  }\href {\doibase 10.1103/PhysRevA.81.053610} {\bibfield  {journal} {\bibinfo
  {journal} {Phys. Rev. A}\ }\textbf {\bibinfo {volume} {81}},\ \bibinfo {eid}
  {053610} (\bibinfo {year} {2010})},\ \Eprint {http://arxiv.org/abs/1002.0869}
  {arXiv:1002.0869 [cond-mat.quant-gas]} \BibitemShut {NoStop}%
\bibitem [{\citenamefont {{Enss}}\ \emph {et~al.}(2011)\citenamefont {{Enss}},
  \citenamefont {{Haussmann}},\ and\ \citenamefont {{Zwerger}}}]{Enss2011}%
  \BibitemOpen
  \bibfield  {author} {\bibinfo {author} {\bibfnamefont {T.}~\bibnamefont
  {{Enss}}}, \bibinfo {author} {\bibfnamefont {R.}~\bibnamefont {{Haussmann}}},
  \ and\ \bibinfo {author} {\bibfnamefont {W.}~\bibnamefont {{Zwerger}}},\
  }\bibfield  {title} {\enquote {\bibinfo {title} {{Viscosity and scale
  invariance in the unitary Fermi gas}},}\ }\href {\doibase
  10.1016/j.aop.2010.10.002} {\bibfield  {journal} {\bibinfo  {journal} {Annals
  of Physics}\ }\textbf {\bibinfo {volume} {326}},\ \bibinfo {pages} {770}
  (\bibinfo {year} {2011})},\ \Eprint {http://arxiv.org/abs/1008.0007}
  {arXiv:1008.0007 [cond-mat.quant-gas]} \BibitemShut {NoStop}%
\bibitem [{\citenamefont {{Kryjevski}}(2014)}]{AK14}%
  \BibitemOpen
  \bibfield  {author} {\bibinfo {author} {\bibfnamefont {A.}~\bibnamefont
  {{Kryjevski}}},\ }\bibfield  {title} {\enquote {\bibinfo {title} {{Shear
  viscosity of the normal phase of a unitary Fermi gas from the ${\varepsilon}$
  expansion}},}\ }\href {\doibase 10.1103/PhysRevA.89.023621} {\bibfield
  {journal} {\bibinfo  {journal} {Phys. Rev. A}\ }\textbf {\bibinfo {volume}
  {89}},\ \bibinfo {pages} {023621} (\bibinfo {year} {2014})},\ \Eprint
  {http://arxiv.org/abs/1206.0059} {arXiv:1206.0059 [cond-mat.quant-gas]}
  \BibitemShut {NoStop}%
\bibitem [{\citenamefont {{Bluhm}}\ and\ \citenamefont
  {{Sch{\"a}fer}}(2016)}]{MBTS16}%
  \BibitemOpen
  \bibfield  {author} {\bibinfo {author} {\bibfnamefont {M.}~\bibnamefont
  {{Bluhm}}}\ and\ \bibinfo {author} {\bibfnamefont {T.}~\bibnamefont
  {{Sch{\"a}fer}}},\ }\bibfield  {title} {\enquote {\bibinfo {title}
  {{Model-Independent Determination of the Shear Viscosity of a Trapped Unitary
  Fermi gas: Application to High-Temperature Data}},}\ }\href {\doibase
  10.1103/PhysRevLett.116.115301} {\bibfield  {journal} {\bibinfo  {journal}
  {Phys. Rev. Lett.}\ }\textbf {\bibinfo {volume} {116}},\ \bibinfo {eid}
  {115301} (\bibinfo {year} {2016})},\ \Eprint
  {http://arxiv.org/abs/1512.00862} {arXiv:1512.00862 [cond-mat.quant-gas]}
  \BibitemShut {NoStop}%
\bibitem [{\citenamefont {{Rameau}}\ \emph {et~al.}(2014)\citenamefont
  {{Rameau}}, \citenamefont {{Reber}}, \citenamefont {{Yang}}, \citenamefont
  {{Akhanjee}}, \citenamefont {{Gu}}, \citenamefont {{Johnson}},\ and\
  \citenamefont {{Campbell}}}]{Johnson14}%
  \BibitemOpen
  \bibfield  {author} {\bibinfo {author} {\bibfnamefont {J.~D.}\ \bibnamefont
  {{Rameau}}}, \bibinfo {author} {\bibfnamefont {T.~J.}\ \bibnamefont
  {{Reber}}}, \bibinfo {author} {\bibfnamefont {H.-B.}\ \bibnamefont {{Yang}}},
  \bibinfo {author} {\bibfnamefont {S.}~\bibnamefont {{Akhanjee}}}, \bibinfo
  {author} {\bibfnamefont {G.~D.}\ \bibnamefont {{Gu}}}, \bibinfo {author}
  {\bibfnamefont {P.~D.}\ \bibnamefont {{Johnson}}}, \ and\ \bibinfo {author}
  {\bibfnamefont {S.}~\bibnamefont {{Campbell}}},\ }\bibfield  {title}
  {\enquote {\bibinfo {title} {{Nearly perfect fluidity in a high-temperature
  superconductor}},}\ }\href {\doibase 10.1103/PhysRevB.90.134509} {\bibfield
  {journal} {\bibinfo  {journal} {Phys. Rev. B}\ }\textbf {\bibinfo {volume}
  {90}},\ \bibinfo {eid} {134509} (\bibinfo {year} {2014})},\ \Eprint
  {http://arxiv.org/abs/1409.5820} {arXiv:1409.5820 [cond-mat.str-el]}
  \BibitemShut {NoStop}%
\bibitem [{\citenamefont {{Halboth}}\ and\ \citenamefont
  {{Metzner}}(2000)}]{CHWM00}%
  \BibitemOpen
  \bibfield  {author} {\bibinfo {author} {\bibfnamefont {C.~J.}\ \bibnamefont
  {{Halboth}}}\ and\ \bibinfo {author} {\bibfnamefont {W.}~\bibnamefont
  {{Metzner}}},\ }\bibfield  {title} {\enquote {\bibinfo {title} {{$d$-Wave
  Superconductivity and Pomeranchuk Instability in the Two-Dimensional Hubbard
  Model}},}\ }\href {\doibase 10.1103/PhysRevLett.85.5162} {\bibfield
  {journal} {\bibinfo  {journal} {Phys. Rev. Lett.}\ }\textbf {\bibinfo
  {volume} {85}},\ \bibinfo {pages} {5162} (\bibinfo {year} {2000})},\ \Eprint
  {http://arxiv.org/abs/cond-mat/0003349} {cond-mat/0003349} \BibitemShut
  {NoStop}%
\bibitem [{\citenamefont {{Oganesyan}}\ \emph {et~al.}(2001)\citenamefont
  {{Oganesyan}}, \citenamefont {{Kivelson}},\ and\ \citenamefont
  {{Fradkin}}}]{OKF01}%
  \BibitemOpen
  \bibfield  {author} {\bibinfo {author} {\bibfnamefont {V.}~\bibnamefont
  {{Oganesyan}}}, \bibinfo {author} {\bibfnamefont {S.~A.}\ \bibnamefont
  {{Kivelson}}}, \ and\ \bibinfo {author} {\bibfnamefont {E.}~\bibnamefont
  {{Fradkin}}},\ }\bibfield  {title} {\enquote {\bibinfo {title} {{Quantum
  theory of a nematic Fermi fluid}},}\ }\href {\doibase
  10.1103/PhysRevB.64.195109} {\bibfield  {journal} {\bibinfo  {journal} {Phys.
  Rev. B}\ }\textbf {\bibinfo {volume} {64}},\ \bibinfo {pages} {195109}
  (\bibinfo {year} {2001})},\ \Eprint {http://arxiv.org/abs/cond-mat/0102093}
  {cond-mat/0102093} \BibitemShut {NoStop}%
\bibitem [{\citenamefont {{Metzner}}\ \emph {et~al.}(2003)\citenamefont
  {{Metzner}}, \citenamefont {{Rohe}},\ and\ \citenamefont
  {{Andergassen}}}]{MRA03}%
  \BibitemOpen
  \bibfield  {author} {\bibinfo {author} {\bibfnamefont {W.}~\bibnamefont
  {{Metzner}}}, \bibinfo {author} {\bibfnamefont {D.}~\bibnamefont {{Rohe}}}, \
  and\ \bibinfo {author} {\bibfnamefont {S.}~\bibnamefont {{Andergassen}}},\
  }\bibfield  {title} {\enquote {\bibinfo {title} {{Soft Fermi Surfaces and
  Breakdown of Fermi-Liquid Behavior}},}\ }\href {\doibase
  10.1103/PhysRevLett.91.066402} {\bibfield  {journal} {\bibinfo  {journal}
  {Phys. Rev. Lett.}\ }\textbf {\bibinfo {volume} {91}},\ \bibinfo {eid}
  {066402} (\bibinfo {year} {2003})},\ \Eprint
  {http://arxiv.org/abs/cond-mat/0303154} {cond-mat/0303154} \BibitemShut
  {NoStop}%
\bibitem [{\citenamefont {Metlitski}\ and\ \citenamefont
  {Sachdev}(2010)}]{Metlitski2010a}%
  \BibitemOpen
  \bibfield  {author} {\bibinfo {author} {\bibfnamefont {M.~A.}\ \bibnamefont
  {Metlitski}}\ and\ \bibinfo {author} {\bibfnamefont {S.}~\bibnamefont
  {Sachdev}},\ }\bibfield  {title} {\enquote {\bibinfo {title} {Quantum phase
  transitions of metals in two spatial dimensions. {I}. {I}sing-nematic
  order},}\ }\href {\doibase 10.1103/PhysRevB.82.075127} {\bibfield  {journal}
  {\bibinfo  {journal} {Phys. Rev. B}\ }\textbf {\bibinfo {volume} {82}},\
  \bibinfo {pages} {075127} (\bibinfo {year} {2010})},\ \Eprint
  {http://arxiv.org/abs/1001.1153} {arXiv:1001.1153 [cond-mat.str-el]}
  \BibitemShut {NoStop}%
\bibitem [{\citenamefont {Dalidovich}\ and\ \citenamefont
  {Lee}(2013)}]{Dalidovich2013}%
  \BibitemOpen
  \bibfield  {author} {\bibinfo {author} {\bibfnamefont {D.}~\bibnamefont
  {Dalidovich}}\ and\ \bibinfo {author} {\bibfnamefont {S.-S.}\ \bibnamefont
  {Lee}},\ }\bibfield  {title} {\enquote {\bibinfo {title} {Perturbative
  non-{F}ermi liquids from dimensional regularization},}\ }\href {\doibase
  10.1103/PhysRevB.88.245106} {\bibfield  {journal} {\bibinfo  {journal} {Phys.
  Rev. B}\ }\textbf {\bibinfo {volume} {88}},\ \bibinfo {pages} {245106}
  (\bibinfo {year} {2013})},\ \Eprint {http://arxiv.org/abs/1307.3170}
  {arXiv:1307.3170 [cond-mat.str-el]} \BibitemShut {NoStop}%
\bibitem [{\citenamefont {{Kovtun}}(2012)}]{LTT}%
  \BibitemOpen
  \bibfield  {author} {\bibinfo {author} {\bibfnamefont {P.}~\bibnamefont
  {{Kovtun}}},\ }\bibfield  {title} {\enquote {\bibinfo {title} {{Lectures on
  hydrodynamic fluctuations in relativistic theories}},}\ }\href {\doibase
  10.1088/1751-8113/45/47/473001} {\bibfield  {journal} {\bibinfo  {journal}
  {J. Phys. A Math. Gen.}\ }\textbf {\bibinfo {volume} {45}},\ \bibinfo {eid}
  {473001} (\bibinfo {year} {2012})},\ \Eprint {http://arxiv.org/abs/1205.5040}
  {arXiv:1205.5040 [hep-th]} \BibitemShut {NoStop}%
\bibitem [{\citenamefont {{Iqbal}}\ and\ \citenamefont
  {{Liu}}(2009)}]{Nabil09}%
  \BibitemOpen
  \bibfield  {author} {\bibinfo {author} {\bibfnamefont {N.}~\bibnamefont
  {{Iqbal}}}\ and\ \bibinfo {author} {\bibfnamefont {H.}~\bibnamefont
  {{Liu}}},\ }\bibfield  {title} {\enquote {\bibinfo {title} {{Universality of
  the hydrodynamic limit in AdS/CFT and the membrane paradigm}},}\ }\href
  {\doibase 10.1103/PhysRevD.79.025023} {\bibfield  {journal} {\bibinfo
  {journal} {\prd}\ }\textbf {\bibinfo {volume} {79}},\ \bibinfo {eid} {025023}
  (\bibinfo {year} {2009})},\ \Eprint {http://arxiv.org/abs/0809.3808}
  {arXiv:0809.3808 [hep-th]} \BibitemShut {NoStop}%
\bibitem [{\citenamefont {{Roychowdhury}}(2015)}]{Raychowdhury15}%
  \BibitemOpen
  \bibfield  {author} {\bibinfo {author} {\bibfnamefont {D.}~\bibnamefont
  {{Roychowdhury}}},\ }\bibfield  {title} {\enquote {\bibinfo {title}
  {{Hydrodynamics from scalar black branes}},}\ }\href {\doibase
  10.1007/JHEP04(2015)162} {\bibfield  {journal} {\bibinfo  {journal} {JHEP}\
  }\textbf {\bibinfo {volume} {4}},\ \bibinfo {eid} {162} (\bibinfo {year}
  {2015})},\ \Eprint {http://arxiv.org/abs/1502.04345} {arXiv:1502.04345
  [hep-th]} \BibitemShut {NoStop}%
\bibitem [{\citenamefont {{Kiritsis}}\ and\ \citenamefont
  {{Matsuo}}(2015)}]{Kiritsis15}%
  \BibitemOpen
  \bibfield  {author} {\bibinfo {author} {\bibfnamefont {E.}~\bibnamefont
  {{Kiritsis}}}\ and\ \bibinfo {author} {\bibfnamefont {Y.}~\bibnamefont
  {{Matsuo}}},\ }\bibfield  {title} {\enquote {\bibinfo {title}
  {{Charge-hyperscaling violating Lifshitz hydrodynamics from black-holes}},}\
  }\href {\doibase 10.1007/JHEP12(2015)076} {\bibfield  {journal} {\bibinfo
  {journal} {JHEP}\ }\textbf {\bibinfo {volume} {12}},\ \bibinfo {eid} {76}
  (\bibinfo {year} {2015})},\ \Eprint {http://arxiv.org/abs/1508.02494}
  {arXiv:1508.02494 [hep-th]} \BibitemShut {NoStop}%
\bibitem [{\citenamefont {{Kuang}}\ and\ \citenamefont
  {{Wu}}(2015)}]{KuangWu15}%
  \BibitemOpen
  \bibfield  {author} {\bibinfo {author} {\bibfnamefont {X.-M.}\ \bibnamefont
  {{Kuang}}}\ and\ \bibinfo {author} {\bibfnamefont {J.-P.}\ \bibnamefont
  {{Wu}}},\ }\bibfield  {title} {\enquote {\bibinfo {title} {{Transport
  coefficients from hyperscaling violating black brane: shear viscosity and
  conductivity}},}\ }\href@noop {} {\bibfield  {journal} {\bibinfo  {journal}
  {ArXiv e-prints}\ } (\bibinfo {year} {2015})},\ \Eprint
  {http://arxiv.org/abs/1511.03008} {arXiv:1511.03008 [hep-th]} \BibitemShut
  {NoStop}%
\bibitem [{\citenamefont {Kolekar}\ \emph {et~al.}(2016)\citenamefont
  {Kolekar}, \citenamefont {Mukherjee},\ and\ \citenamefont {Narayan}}]{KMN16}%
  \BibitemOpen
  \bibfield  {author} {\bibinfo {author} {\bibfnamefont {K.~S.}\ \bibnamefont
  {Kolekar}}, \bibinfo {author} {\bibfnamefont {D.}~\bibnamefont {Mukherjee}},
  \ and\ \bibinfo {author} {\bibfnamefont {K.}~\bibnamefont {Narayan}},\
  }\bibfield  {title} {\enquote {\bibinfo {title} {{Hyperscaling violation and
  the shear diffusion constant}},}\ }\href {\doibase
  10.1016/j.physletb.2016.06.046} {\bibfield  {journal} {\bibinfo  {journal}
  {Phys. Lett.}\ }\textbf {\bibinfo {volume} {B760}},\ \bibinfo {pages} {86}
  (\bibinfo {year} {2016})},\ \Eprint {http://arxiv.org/abs/1604.05092}
  {arXiv:1604.05092 [hep-th]} \BibitemShut {NoStop}%
\bibitem [{\citenamefont {{Liu}}\ \emph {et~al.}(2011)\citenamefont {{Liu}},
  \citenamefont {{McGreevy}},\ and\ \citenamefont {{Vegh}}}]{Liu09}%
  \BibitemOpen
  \bibfield  {author} {\bibinfo {author} {\bibfnamefont {H.}~\bibnamefont
  {{Liu}}}, \bibinfo {author} {\bibfnamefont {J.}~\bibnamefont {{McGreevy}}}, \
  and\ \bibinfo {author} {\bibfnamefont {D.}~\bibnamefont {{Vegh}}},\
  }\bibfield  {title} {\enquote {\bibinfo {title} {{Non-Fermi liquids from
  holography}},}\ }\href {\doibase 10.1103/PhysRevD.83.065029} {\bibfield
  {journal} {\bibinfo  {journal} {Phys. Rev. D}\ }\textbf {\bibinfo {volume}
  {83}},\ \bibinfo {eid} {065029} (\bibinfo {year} {2011})},\ \Eprint
  {http://arxiv.org/abs/0903.2477} {arXiv:0903.2477 [hep-th]} \BibitemShut
  {NoStop}%
\bibitem [{\citenamefont {{{\v C}ubrovi{\'c}}}\ \emph
  {et~al.}(2009)\citenamefont {{{\v C}ubrovi{\'c}}}, \citenamefont {{Zaanen}},\
  and\ \citenamefont {{Schalm}}}]{Zaanen09}%
  \BibitemOpen
  \bibfield  {author} {\bibinfo {author} {\bibfnamefont {M.}~\bibnamefont {{{\v
  C}ubrovi{\'c}}}}, \bibinfo {author} {\bibfnamefont {J.}~\bibnamefont
  {{Zaanen}}}, \ and\ \bibinfo {author} {\bibfnamefont {K.}~\bibnamefont
  {{Schalm}}},\ }\bibfield  {title} {\enquote {\bibinfo {title} {{String
  Theory, Quantum Phase Transitions, and the Emergent Fermi Liquid}},}\ }\href
  {\doibase 10.1126/science.1174962} {\bibfield  {journal} {\bibinfo  {journal}
  {Science}\ }\textbf {\bibinfo {volume} {325}},\ \bibinfo {pages} {439}
  (\bibinfo {year} {2009})},\ \Eprint {http://arxiv.org/abs/0904.1993}
  {arXiv:0904.1993 [hep-th]} \BibitemShut {NoStop}%
\bibitem [{\citenamefont {Charmousis}\ \emph {et~al.}(2010)\citenamefont
  {Charmousis}, \citenamefont {Gouteraux}, \citenamefont {Kim}, \citenamefont
  {Kiritsis},\ and\ \citenamefont {Meyer}}]{Kiritsis10}%
  \BibitemOpen
  \bibfield  {author} {\bibinfo {author} {\bibfnamefont {C.}~\bibnamefont
  {Charmousis}}, \bibinfo {author} {\bibfnamefont {B.}~\bibnamefont
  {Gouteraux}}, \bibinfo {author} {\bibfnamefont {B.~S.}\ \bibnamefont {Kim}},
  \bibinfo {author} {\bibfnamefont {E.}~\bibnamefont {Kiritsis}}, \ and\
  \bibinfo {author} {\bibfnamefont {R.}~\bibnamefont {Meyer}},\ }\bibfield
  {title} {\enquote {\bibinfo {title} {{Effective Holographic Theories for
  low-temperature condensed matter systems}},}\ }\href {\doibase
  10.1007/JHEP11(2010)151} {\bibfield  {journal} {\bibinfo  {journal} {JHEP}\
  }\textbf {\bibinfo {volume} {11}},\ \bibinfo {pages} {151} (\bibinfo {year}
  {2010})},\ \Eprint {http://arxiv.org/abs/1005.4690} {arXiv:1005.4690
  [hep-th]} \BibitemShut {NoStop}%
\bibitem [{\citenamefont {Goldstein}\ \emph {et~al.}(2010)\citenamefont
  {Goldstein}, \citenamefont {Kachru}, \citenamefont {Prakash},\ and\
  \citenamefont {Trivedi}}]{Trivedi10}%
  \BibitemOpen
  \bibfield  {author} {\bibinfo {author} {\bibfnamefont {K.}~\bibnamefont
  {Goldstein}}, \bibinfo {author} {\bibfnamefont {S.}~\bibnamefont {Kachru}},
  \bibinfo {author} {\bibfnamefont {S.}~\bibnamefont {Prakash}}, \ and\
  \bibinfo {author} {\bibfnamefont {S.~P.}\ \bibnamefont {Trivedi}},\
  }\bibfield  {title} {\enquote {\bibinfo {title} {{Holography of Charged
  Dilaton Black Holes}},}\ }\href {\doibase 10.1007/JHEP08(2010)078} {\bibfield
   {journal} {\bibinfo  {journal} {JHEP}\ }\textbf {\bibinfo {volume} {08}},\
  \bibinfo {pages} {078} (\bibinfo {year} {2010})},\ \Eprint
  {http://arxiv.org/abs/0911.3586} {arXiv:0911.3586 [hep-th]} \BibitemShut
  {NoStop}%
\bibitem [{\citenamefont {Ogawa}\ \emph {et~al.}(2012)\citenamefont {Ogawa},
  \citenamefont {Takayanagi},\ and\ \citenamefont {Ugajin}}]{Ogawa11}%
  \BibitemOpen
  \bibfield  {author} {\bibinfo {author} {\bibfnamefont {N.}~\bibnamefont
  {Ogawa}}, \bibinfo {author} {\bibfnamefont {T.}~\bibnamefont {Takayanagi}}, \
  and\ \bibinfo {author} {\bibfnamefont {T.}~\bibnamefont {Ugajin}},\
  }\bibfield  {title} {\enquote {\bibinfo {title} {{Holographic Fermi Surfaces
  and Entanglement Entropy}},}\ }\href {\doibase 10.1007/JHEP01(2012)125}
  {\bibfield  {journal} {\bibinfo  {journal} {JHEP}\ }\textbf {\bibinfo
  {volume} {01}},\ \bibinfo {pages} {125} (\bibinfo {year} {2012})},\ \Eprint
  {http://arxiv.org/abs/1111.1023} {arXiv:1111.1023 [hep-th]} \BibitemShut
  {NoStop}%
\bibitem [{\citenamefont {{Huijse}}\ \emph {et~al.}(2012)\citenamefont
  {{Huijse}}, \citenamefont {{Sachdev}},\ and\ \citenamefont
  {{Swingle}}}]{HSS11}%
  \BibitemOpen
  \bibfield  {author} {\bibinfo {author} {\bibfnamefont {L.}~\bibnamefont
  {{Huijse}}}, \bibinfo {author} {\bibfnamefont {S.}~\bibnamefont {{Sachdev}}},
  \ and\ \bibinfo {author} {\bibfnamefont {B.}~\bibnamefont {{Swingle}}},\
  }\bibfield  {title} {\enquote {\bibinfo {title} {{Hidden Fermi surfaces in
  compressible states of gauge-gravity duality}},}\ }\href {\doibase
  10.1103/PhysRevB.85.035121} {\bibfield  {journal} {\bibinfo  {journal}
  {\prb}\ }\textbf {\bibinfo {volume} {85}},\ \bibinfo {eid} {035121} (\bibinfo
  {year} {2012})},\ \Eprint {http://arxiv.org/abs/1112.0573} {arXiv:1112.0573
  [cond-mat.str-el]} \BibitemShut {NoStop}%
\bibitem [{\citenamefont {Polchinski}\ and\ \citenamefont
  {Silverstein}(2012)}]{Polchinski:2012nh}%
  \BibitemOpen
  \bibfield  {author} {\bibinfo {author} {\bibfnamefont {J.}~\bibnamefont
  {Polchinski}}\ and\ \bibinfo {author} {\bibfnamefont {E.}~\bibnamefont
  {Silverstein}},\ }\bibfield  {title} {\enquote {\bibinfo {title}
  {{Large-density field theory, viscosity, and `$2k_F$' singularities from
  string duals}},}\ }\href {\doibase 10.1088/0264-9381/29/19/194008} {\bibfield
   {journal} {\bibinfo  {journal} {Class. Quant. Grav.}\ }\textbf {\bibinfo
  {volume} {29}},\ \bibinfo {pages} {194008} (\bibinfo {year} {2012})},\
  \Eprint {http://arxiv.org/abs/1203.1015} {arXiv:1203.1015 [hep-th]}
  \BibitemShut {NoStop}%
\bibitem [{\citenamefont {Faulkner}\ and\ \citenamefont
  {Iqbal}(2013)}]{Faulkner:2012gt}%
  \BibitemOpen
  \bibfield  {author} {\bibinfo {author} {\bibfnamefont {T.}~\bibnamefont
  {Faulkner}}\ and\ \bibinfo {author} {\bibfnamefont {N.}~\bibnamefont
  {Iqbal}},\ }\bibfield  {title} {\enquote {\bibinfo {title} {{Friedel
  oscillations and horizon charge in 1D holographic liquids}},}\ }\href
  {\doibase 10.1007/JHEP07(2013)060} {\bibfield  {journal} {\bibinfo  {journal}
  {JHEP}\ }\textbf {\bibinfo {volume} {07}},\ \bibinfo {pages} {060} (\bibinfo
  {year} {2013})},\ \Eprint {http://arxiv.org/abs/1207.4208} {arXiv:1207.4208
  [hep-th]} \BibitemShut {NoStop}%
\bibitem [{\citenamefont {Sachdev}(2012)}]{Sachdev:2012tj}%
  \BibitemOpen
  \bibfield  {author} {\bibinfo {author} {\bibfnamefont {S.}~\bibnamefont
  {Sachdev}},\ }\bibfield  {title} {\enquote {\bibinfo {title} {{Compressible
  quantum phases from conformal field theories in 2+1 dimensions}},}\ }\href
  {\doibase 10.1103/PhysRevD.86.126003} {\bibfield  {journal} {\bibinfo
  {journal} {Phys. Rev. D}\ }\textbf {\bibinfo {volume} {86}},\ \bibinfo
  {pages} {126003} (\bibinfo {year} {2012})},\ \Eprint
  {http://arxiv.org/abs/1209.1637} {arXiv:1209.1637 [hep-th]} \BibitemShut
  {NoStop}%
\bibitem [{\citenamefont {Eberlein}\ \emph {et~al.}(2016)\citenamefont
  {Eberlein}, \citenamefont {Mandal},\ and\ \citenamefont
  {Sachdev}}]{Eberlein2016}%
  \BibitemOpen
  \bibfield  {author} {\bibinfo {author} {\bibfnamefont {A.}~\bibnamefont
  {Eberlein}}, \bibinfo {author} {\bibfnamefont {I.}~\bibnamefont {Mandal}}, \
  and\ \bibinfo {author} {\bibfnamefont {S.}~\bibnamefont {Sachdev}},\
  }\bibfield  {title} {\enquote {\bibinfo {title} {{Hyperscaling violation at
  the Ising-nematic quantum critical point in two-dimensional metals}},}\
  }\href {\doibase 10.1103/PhysRevB.94.045133} {\bibfield  {journal} {\bibinfo
  {journal} {Phys. Rev. B}\ }\textbf {\bibinfo {volume} {94}},\ \bibinfo
  {pages} {045133} (\bibinfo {year} {2016})},\ \Eprint
  {http://arxiv.org/abs/1605.00657} {arXiv:1605.00657 [cond-mat.str-el]}
  \BibitemShut {NoStop}%
\bibitem [{\citenamefont {Schwartz}(2014)}]{schwartz2014quantum}%
  \BibitemOpen
  \bibfield  {author} {\bibinfo {author} {\bibfnamefont {M.~D.}\ \bibnamefont
  {Schwartz}},\ }\href@noop {} {\emph {\bibinfo {title} {Quantum field theory
  and the standard model}}}\ (\bibinfo  {publisher} {Cambridge University
  Press},\ \bibinfo {year} {2014})\BibitemShut {NoStop}%
\bibitem [{\citenamefont {{Fritz}}(2011)}]{fritz11}%
  \BibitemOpen
  \bibfield  {author} {\bibinfo {author} {\bibfnamefont {L.}~\bibnamefont
  {{Fritz}}},\ }\bibfield  {title} {\enquote {\bibinfo {title} {{Quantum
  critical transport at a semimetal-to-insulator transition on the honeycomb
  lattice}},}\ }\href {\doibase 10.1103/PhysRevB.83.035125} {\bibfield
  {journal} {\bibinfo  {journal} {\prb}\ }\textbf {\bibinfo {volume} {83}},\
  \bibinfo {eid} {035125} (\bibinfo {year} {2011})},\ \Eprint
  {http://arxiv.org/abs/1012.0263} {arXiv:1012.0263 [cond-mat.str-el]}
  \BibitemShut {NoStop}%
\bibitem [{\citenamefont {Kamenev}(2011)}]{Kamenev2011}%
  \BibitemOpen
  \bibfield  {author} {\bibinfo {author} {\bibfnamefont {A.}~\bibnamefont
  {Kamenev}},\ }\href
  {http://www.cambridge.org/us/academic/subjects/physics/condensed-matter-physics-nanoscience-and-mesoscopic-physics/field-theory-non-equilibrium-systems}
  {\emph {\bibinfo {title} {{Field Theory of Non-Equilibrium Systems}}}}\
  (\bibinfo  {publisher} {Cambridge University Press, Cambridge, UK},\ \bibinfo
  {year} {2011})\BibitemShut {NoStop}%
\bibitem [{\citenamefont {{Patel}}\ \emph {et~al.}(2015)\citenamefont
  {{Patel}}, \citenamefont {{Strack}},\ and\ \citenamefont
  {{Sachdev}}}]{Patel2015}%
  \BibitemOpen
  \bibfield  {author} {\bibinfo {author} {\bibfnamefont {A.~A.}\ \bibnamefont
  {{Patel}}}, \bibinfo {author} {\bibfnamefont {P.}~\bibnamefont {{Strack}}}, \
  and\ \bibinfo {author} {\bibfnamefont {S.}~\bibnamefont {{Sachdev}}},\
  }\bibfield  {title} {\enquote {\bibinfo {title} {{Hyperscaling at the spin
  density wave quantum critical point in two-dimensional metals}},}\ }\href
  {\doibase 10.1103/PhysRevB.92.165105} {\bibfield  {journal} {\bibinfo
  {journal} {Phys. Rev. B}\ }\textbf {\bibinfo {volume} {92}},\ \bibinfo {eid}
  {165105} (\bibinfo {year} {2015})},\ \Eprint
  {http://arxiv.org/abs/1507.05962} {arXiv:1507.05962 [cond-mat.str-el]}
  \BibitemShut {NoStop}%
\bibitem [{\citenamefont {{Maslov}}\ \emph {et~al.}(2011)\citenamefont
  {{Maslov}}, \citenamefont {{Yudson}},\ and\ \citenamefont
  {{Chubukov}}}]{Maslov2011}%
  \BibitemOpen
  \bibfield  {author} {\bibinfo {author} {\bibfnamefont {D.~L.}\ \bibnamefont
  {{Maslov}}}, \bibinfo {author} {\bibfnamefont {V.~I.}\ \bibnamefont
  {{Yudson}}}, \ and\ \bibinfo {author} {\bibfnamefont {A.~V.}\ \bibnamefont
  {{Chubukov}}},\ }\bibfield  {title} {\enquote {\bibinfo {title} {{Resistivity
  of a Non-Galilean-Invariant Fermi Liquid near Pomeranchuk Quantum
  Criticality}},}\ }\href {\doibase 10.1103/PhysRevLett.106.106403} {\bibfield
  {journal} {\bibinfo  {journal} {Phys. Rev. Lett.}\ }\textbf {\bibinfo
  {volume} {106}},\ \bibinfo {eid} {106403} (\bibinfo {year} {2011})},\ \Eprint
  {http://arxiv.org/abs/1012.0069} {arXiv:1012.0069 [cond-mat.str-el]}
  \BibitemShut {NoStop}%
\bibitem [{\citenamefont {Negele}\ and\ \citenamefont
  {Orland}(1998)}]{Negele1998}%
  \BibitemOpen
  \bibfield  {author} {\bibinfo {author} {\bibfnamefont {J.~W.}\ \bibnamefont
  {Negele}}\ and\ \bibinfo {author} {\bibfnamefont {H.}~\bibnamefont
  {Orland}},\ }\href
  {https://westviewpress.com/books/quantum-many-particle-systems/} {\emph
  {\bibinfo {title} {{Quantum {M}any-{P}article {S}ystems}}}},\ edited by\
  \bibinfo {editor} {\bibfnamefont {D.}~\bibnamefont {Pines}},\ Advanced Book
  Classics\ (\bibinfo  {publisher} {Westview Press, Reading, Massachusetts},\
  \bibinfo {year} {1998})\BibitemShut {NoStop}%
\end{thebibliography}%

\end{document}